\documentclass[epj]{svjour}
\usepackage{amsmath}
\usepackage{amssymb}
\usepackage{graphicx}
\usepackage[T1]{fontenc}
\usepackage{slashed}
\usepackage{subcaption}
\usepackage{bbm}
\usepackage{tabularx}
\usepackage{physics}
\usepackage[table]{xcolor}
\usepackage{multirow}
\usepackage{array} 
\usepackage{cuted}
\usepackage{upgreek}
\usepackage{adjustbox}

\renewcommand{\Gamma}{\Upgamma}

\definecolor{mycolor}{rgb}{0.8, 0.4, 0.4}
\newcommand{\x}{\cellcolor{yellow!20}}
\newcommand{\y}{\cellcolor{red!20}}
\newcommand{\z}{\cellcolor{blue!20}}
\begin{document}
\title{Symmetric Tensor Coupling in Holographic Mean-Field Theory: Deformed Dirac Cones}
\author{Moongul Byun\inst{1}\thanks{\email{moongulbyun@hanyang.ac.kr}} \and Taewon Yuk\inst{1}\thanks{\email{tae1yuk@gmail.com}} \and Sang-Jin Sin\inst{1}\thanks{\email{sangjin.sin@gmail.com} (corresponding author)}
%
}                     
%

%
\institute{Department of Physics, Hanyang University, Seoul 04763, South Korea}
\date{Received: date / Revised version: date}
%
\abstract{
We extend the holographic mean-field theory to rank-two symmetric tensor field as an external source coupled with fermion. 
We classify the roles of symmetric tensor coupling according to the effect on the spectral density: cone-angle change, squashing, and tilting of the spectral light cones.
The over-tilted light cone is also achieved in a generalized prescription, which consistently retains the causality condition.
Our results provide agreements between the holographic spectra with those observed in real materials, such as type-II Dirac cones and strained graphene.
\PACS{
      {PACS-key}{discribing text of that key}   \and
      {PACS-key}{discribing text of that key}
     } 
} 
\maketitle
\section{Introduction}
\label{section1}

The AdS/CFT correspondence has offered a powerful way to investigate strongly coupled quantum systems by examining their dual classical gravity dual \cite{Maldacena:1997re,Witten:1998qj}. 
Early studies of holography provided the set up to calculate the retarded Green’s functions and revealed the basic aspects of dynamics in boundary theories. 
Subsequently, the method has been widely used in condensed matter systems, most notably in 
transports \cite{Hartnoll:2009sz,Faulkner:2009wj,Faulkner:2010da,Donos:2013eha,Blake:2013bqa,Donos:2014yya,Kim:2014bza,Davison_2014,Blake_2015,Hartnoll:2016apf,Ge:2016lyn,Donos:2018kkm} and  holographic superconductors \cite{Hartnoll:2008kx,Hartnoll:2008vx,Gubser:2008px,Faulkner:2009am,Horowitz_2009,Horowitz_2011,Kim:2013oba} as well as fermion dynamics \cite{Iqbal:2009fd,Laia:2011zn,Faulkner:2013bna}.

Describing such strongly correlated system via gravity dual, we assume the existence of an approximate dual description for such strongly interacting systems because it is hard to find the exact gravity dual of the given system.
Motivated by this perspective, the holographic mean-field theory (HMFT) was developed systematically and has been widely accepted as an effective method to incorporate the basic features of the fermion spectrum \cite{Oh:2020cym,Sukrakarn:2023ncp} appearing in the ARPES data of real fermion.
This mean field approach serves as a representative theory for the purpose of studying various types of the gaps and condensations as well as the singularity types of the Green functions for the strongly interacting systems.
Furthermore, because the holographic theory as a continuum field theory does not encode the condensed matter system’s detail, it is useful to study all possible types of interactions together and classify their spectral behavior to match the physical system’s spectral pattern. 
In \cite{Oh:2020cym} we put forwarded a step to such direction by introducing order parameters $B_{A_{1}A_{2}A_{3}\cdots}$ that break symmetries in the bulk spacetime through fermionic bilinear couplings of the form
\begin{equation}
	\mathcal{L}_{\mathrm{int}} = B_{A_{1}A_{2}A_{3}\cdots}\bar{\psi}\Gamma^{A_{1}A_{2}A_{3}\cdots}\psi,
\end{equation}
representing various types of condensates in the holographic setup analogous to fermion bilinear coupled to the Hubbard-Stratonovich field in the conventional mean field theory.
It was observed that the holographic Green's functions from the mean-field theory contains many of the features in the spectral functions of the fermions in the real system so that it was suggested that \cite{Sukrakarn:2023ncp} the order parameter field due to the specific symmetry breaking could be used as a way to encode the effect of certain  lattices. Naturally such scheme was applied to various types of holographic superconductors \cite{Yuk:2022lof,Ghorai:2023wpu,PhysRevD.109.066004} as well as to the Kondo lattice physics \cite{Im:2023ffg,Han:2024rbr}.

In \cite{Oh:2020cym,Sukrakarn:2023ncp}, scalar, vector, and antisymmetric-tensor couplings were considered and their spectral effects were classified, whereas the symmetric-tensor coupling,
\begin{equation}
	\mathcal{L}_{\mathrm{int}} = \Phi_{MN}\bar{\psi}\Gamma^{(M}D^{N)}\psi,
\end{equation}
with $\Phi_{MN}$ being the rank-2 symmetric tensor field, was not considered. 
Also previous literature on symmetric tensors primarily focused on purely spatial components ($\Phi_{ii}$ and $\Phi_{ij}$), identifying them with $d$-wave order parameters in holographic superconductors with  gap features in spectral densities and $U(1)$-symmetry breaking are present \cite{Horowitz_2009,Horowitz_2011,Kim:2013oba,PhysRevD.109.066004}. 
While the tensor field as the $d$-wave order parameters is considered as a complex field, our primary focus lies on the case where $\Phi_{MN}$ is a real one.
Traceless spatial symmetric tensors were also studied and associated with quantum spin nematic phases \cite{PhysRevB.64.195109,PhysRevB.96.235140}, realized through linearized gravity \cite{PhysRevB.109.L220407}. 
However, the effects of time-time ($\Phi_{tt}$), time-space ($\Phi_{ti}$) have not been considered and general correspondence of the symmetric tensor and their counterparts in condensed matter systems have not been identified. 

This work seeks to bridge the gap in understanding by systematically analyzing the effects of all tensor components from the view of the holographic mean-field theory (HMFT) framework: 
we extend HMFT by coupling one- and two-flavor spinors to a rank-two symmetric tensor in $\mathrm{AdS}_{5}$, providing analytic evaluations of Green’s functions and spectral densities (SDs)
at the level of the probe without full backreaction. 

It turns out that boundary components of the symmetric tensor field $\Phi_{\mu\nu}$ coupled to Dirac spinors induce three distinct types of deformations in the resulting spectral densities: \textit{cone-angle change}, \textit{squashing}, and \textit{tilting}. 
We present these possibilities schematically as follows:
\begin{equation}
	\begin{adjustbox}{max width=\columnwidth}
		$\Phi_{\mu\nu} =
		\begin{pmatrix}
			\x \Phi_{tt}  & \z \Phi_{tx}  & \z \Phi_{ty} & \z \Phi_{tz}\\
			\z \Phi_{tx}  & \y \Phi_{xx}  & \y \Phi_{xy} & \y \Phi_{xz}\\
			\z \Phi_{ty}  & \y \Phi_{xy}  & \y \Phi_{yy} & \y \Phi_{yz}\\
			\z \Phi_{tz}  & \y \Phi_{xz}  & \y \Phi_{yz} & \y \Phi_{zz}\\
		\end{pmatrix}
		\quad\text{where}\quad\left\{\begin{aligned}
			&\fcolorbox{black}{yellow!20}{\rule{0pt}{6pt}\rule{6pt}{0pt}}:\, \text{Cone-angle change},\\
			&\fcolorbox{black}{red!20}{\rule{0pt}{6pt}\rule{6pt}{0pt}}:\, \text{Squashing},\\
			&\fcolorbox{black}{blue!20}{\rule{0pt}{6pt}\rule{6pt}{0pt}}:\, \text{Tilting} + \text{Squashing}.
		\end{aligned}\right.
		$
	\end{adjustbox}
\end{equation}
These symmetric tensor couplings can therefore reshape the spectral densities in various ways, with each   component of $\Phi_{\mu\nu}$ serving as a parameter that   modifies Fermi velocities or induces tilts in specific direction.

It is remarkable that each symmetric tensor component can find a corresponding real system parameter counterpart in condensed matter systems. The strength of $\Phi_{\mu\nu}$ can be associated with specific parameters in condensed matter systems by analyzing the spectral features under the symmetric tensor coupling.

For instance, the $\Phi_{tt}$ component induces cone-angle change of the spectral densities, leading to the modification of the Fermi velocity under a uniform electric field \cite{D_az_Fern_ndez_2017}.
Also, critically decreased Fermi velocity due to strong $\Phi_{tt}$ coupling can be associated with the local flat band near the Dirac point observed in twisted bilayer graphene at the magic angle \cite{Lisi_2020}.

In the case of $\Phi_{ti}$, the coupling induces a tilting 
of the SDs, which can be interpreted as a tilt parameter, leading to  type-I, type-II, or type-III Dirac cones \cite{PhysRevX.9.031010} according to the coupling strength. For  two flavour system,  such classification leads to the corresponding Weyl semimetals \cite{Soluyanov_2015} of type-I, -II, -III.
It turns out that  such tilts can be realized through metric deformations and modifications of the Lorentz algebra \cite{Volovik_2016,Volovik_2021,PhysRevResearch.4.033237,PhysRevB.100.045144,Seo2026}.

On the other hand, the spatial components $\Phi_{ii}$ and $\Phi_{ij}$ induce squashing, i.e., anisotropic scaling, on spectral density in momentum space.
The spectral deformations induced by these spatial components are analogous to those associated with rank-2 quadrupolar order parameters in quantum spin-nematic phases which also give rise to elliptical Fermi surfaces \cite{PhysRevB.64.195109,PhysRevB.96.235140}.
Indeed, the effects of $\Phi_{ij}$ resemble lattice deformations, such as strain and shear strain, described by a rank-2 strain tensor and observed in systems like strained and twisted (bilayer) graphene \cite{PhysRevB.88.085430,PhysRevResearch.4.L022027}, as well as in $\mathrm{Bi}_{2}\mathrm{Se}_{3}$-class materials \cite{Brems_2018,PhysRevB.84.195425}.

\begin{figure}[tb]
	\centering
	\includegraphics[width =0.75\linewidth]{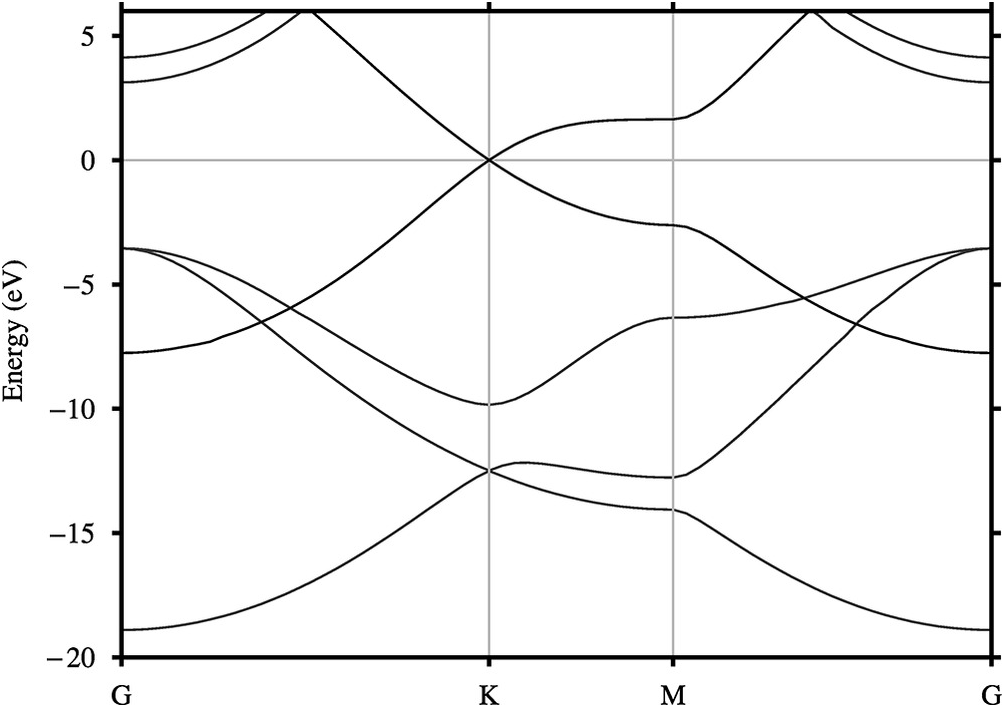}
	\caption{Graphene band structure, adapted from \cite{Katsnelson_2020}. Lattice deformations or interactions can reshape the Dirac cones in ways that may be effectively described by and identified with the effects of symmetric tensor couplings $\Phi_{\mu\nu}$ on spectral densities.}
	\label{figure1}
\end{figure}

We summarize the correspondence between the symmetric tensor coupling and the real material examples in condensed matter in the table~\ref{table1}.
\begin{table*}
	\centering\footnotesize
	\renewcommand{\arraystretch}{1.2} 
	\resizebox{1.0\textwidth}{!}{
		\begin{tabular}{|>{\centering\arraybackslash}m{0.13\textwidth}|
				>{\centering\arraybackslash}m{0.19\textwidth}|
				>{}m{0.6\textwidth}|
				>{\centering\arraybackslash}m{0.15\textwidth}|}
			\hline
			\begin{minipage}[c]{\linewidth}\centering\vspace{3pt}
				\textbf{Tensor Components}\vspace{3pt}
			\end{minipage} 
			& \begin{minipage}[c]{\linewidth}\centering\vspace{3pt}
				\textbf{Features in Holographic SD}\vspace{3pt}
			\end{minipage} 
			& \begin{minipage}[c]{\linewidth}\centering\vspace{3pt}
				\textbf{Phenomena in Real Systems}\vspace{3pt}
			\end{minipage} 
			& \begin{minipage}[c]{\linewidth}\centering\vspace{3pt}
				\textbf{System}\\\textbf{Parameter}\vspace{3pt}
			\end{minipage} \\ 
			\hline
			\multirow{2}{*}{\begin{minipage}[c]{\linewidth}\centering
					\large{ $\Phi_{tt}$} 
			\end{minipage}}
			& \multirow{2}{*}{\begin{minipage}[c]{\linewidth}\centering
					Cone-angle change 
			\end{minipage}}
			& Tuning Fermi velocity in Dirac materials via a uniform electric field \cite{D_az_Fern_ndez_2017}.
			& Electric field \\ \cline{3-4}
			& & Flat bands at local Dirac points emerge in twisted bilayer graphene at the magic angle \cite{Lisi_2020}.
			& Twist angle \\
			\hline
			\large{ $\Phi_{ti}$} 
			& \begin{minipage}[c]{\linewidth}\centering
				Tilting and squashing 
			\end{minipage}
			& Rotated energy dispersion in tilted type-I, type-II, and type-III Dirac/Weyl cones \cite{PhysRevX.9.031010}.
			& Tilt parameter \\ 
			\hline
			\multirow{4}{*}{\begin{minipage}[c]{\linewidth}\centering
					\large{ $\Phi_{ii}$, $\Phi_{ij}$}
			\end{minipage}}
			& \multirow{4}{*}{\begin{minipage}[c]{\linewidth}\centering
					Squashing and rotation\\ 
			\end{minipage}}
			& Spatial components of symmetric tensor couplings identified as $d$-wave order parameters for holographic superconductors if there is a gap feature \cite{Kim:2013oba,PhysRevD.109.066004}.
			& $d$-wave order parameters \\ \cline{3-4}
			& & Quadrupolar two-rank tensors for quantum spin nematic phases \cite{PhysRevB.64.195109,PhysRevB.96.235140}, which can be realized by linearized gravity \cite{PhysRevB.109.L220407}.
			& Nematic order parameter \\ \cline{3-4}
			& & Lattice deformation (strain and shear strain) in (bilayer) graphene \cite{PhysRevB.88.085430,PhysRevResearch.4.L022027} and $\mathrm{Bi}_{2}\mathrm{Se}_{3}$ \cite{Brems_2018}.
			& Strain tensor \\ \cline{3-4}
			& & Anisotropic deformation of Dirac cones of superlattice graphene due to a periodic potential \cite{Park_2008}.
			& Superlattice structure \\ \hline
		\end{tabular}
	}
	\caption{Summary: Classification of symmetric tensor coupling components, their spectral features and  examples in condensed matter systems.}
	\label{table1}
\end{table*}
Notice that \textit{the $\Phi_{\mu\nu}$ components can be identified by matching measured Fermi velocities, cone angles, and tilting angles of the local energy dispersion for a given material}, which constitute our primary results.
For instance, figure~\ref{figure1} shows the band structure of graphene \cite{Katsnelson_2020} as an illustrative example of a Dirac dispersion that could be both qualitatively and quantitatively associated with symmetric tensor couplings at each local crossing point.
\vskip 0.5cm
The rest of this paper is organized as follows.
In section~\ref{section2}, we define the bulk and boundary actions, introduce the symmetric tensor coupling, and derive the Dirac equation for holographic fermions.
Sections~\ref{section3} to \ref{section4} derive analytic retarded Green’s functions for one- and two-flavor spinors, categorizing the resulting spectral densities based on their scaling and rotational behavior under each $\Phi_{\mu\nu}$ component.
We also propose a formalism to obtain well-defined over-tilted spectral density, which shows the positive-definite its value, in section~\ref{section3}.
In section~\ref{section6}, we explore potential applications of the symmetric tensor coupling by comparing its components with analogous parameters in condensed matter systems.
Finally, section~\ref{section7} summarizes the main findings and discusses potential applications of this holographic mean-field theory approach to other condensed matter systems.

\section{Holographic fermions  with symmetric tensor coupling}
\label{section2}
We set up the notations for  holographic fermions. 
Throughout this paper,     $M, N, \cdots$ or $t, x, \cdots$ represent   bulk indices on a five-dimensional   manifold $\mathcal{M}$, while $a, b, \cdots$ or $\underline{t}, \underline{x}, \cdots$ denote  the local tangent space indices in the vielbein formalism.
Additionally, the indices $\mu, \nu, \cdots$ represent   boundary   indices on a four-dimensional   $\partial\mathcal{M}$, while $i, j, \cdots$  denotes  the spatial indices at the   boundary.
Now we define the action by 
\begin{equation}
	\label{eq2.1}
	S_{\mathrm{tot}} = S_{\mathrm{bulk}} + S_{\mathrm{bdy}} + S_{g} + S_{\mathrm{int}},
\end{equation}
\begin{equation}
	\label{eq2.2}
	S_{\mathrm{bulk}} = i\int_{\mathcal{M}} d^{5}x\sqrt{-g}\,\bar{\psi}(\overleftrightarrow{\slashed{D}} - m)\psi,
\end{equation}
\begin{equation}
	\label{eq2.3}
	S_{\mathrm{bdy}} = -i\int_{\partial\mathcal{M}}d^{4}x\sqrt{-gg^{rr}}\bar{\psi}\psi,
\end{equation}
\begin{equation}
	\label{eq2.4}
	S_{g} = \int_{\mathcal{M}}d^{5}x\sqrt{-g}\left(R - 2\Lambda\right),
\end{equation}
\begin{equation}
	\label{eq2.5}
	S_{\mathrm{int}} = i\int_{\mathcal{M}}d^{5}x\sqrt{-g}\bar{\psi}\Phi_{MN}\Gamma^{(M}\overleftrightarrow{D}^{N)}\psi.
\end{equation}
Here, $\overleftrightarrow{D} = \frac{1}{2}(\overrightarrow{D} - \overleftarrow{D})$, while $\slashed{D} = \Gamma^{M}D_{M} = \Gamma^{M}(\partial_{M} + \frac{1}{4}\omega_{abM}\Gamma^{ab})$, and $\omega_{abM}$ denotes the spin connection. 
The field $g_{MN}$ is the background metric tensor associated with the cosmological constant $\Lambda$, whereas $\Phi_{MN}$ represents a five-dimensional symmetric rank-2 tensor.
In \eqref{eq2.3}, $g^{rr}$ denotes the radial-radial component of the inverse metric.
In this paper, we consider fermions in the probe limit. 

We introduce the rank-2 field  $\Phi_{MN}(r)$ in the form of a symmetric tensor, as given in \eqref{eq2.5}, with the indices $M, N = t, x, y, z, r$. 
Throughout this paper, we particularly focus on the boundary components of $\Phi_{MN}(r)$, defined as
\begin{equation}
	\label{eq2.7}
	\Phi_{\mu\nu}(r) := \left.\Phi_{MN}(r)\right|_{M, N \, = \, t, x, y, z}.
\end{equation}
For simplicity, we occasionally denote this as $\Phi_{\mu\nu}$. 

The basic motivation for the interaction in \eqref{eq2.5} is that it provides a minimal coupling between the fermionic stress-energy tensor $T_{\mu\nu}$ and the spin-two field $\Phi_{\mu\nu}$, coupled in the form of $T_{\mu\nu}\Phi^{\mu\nu}$.
This coupling therefore allows us to probe how the spin-two field modifies the fermion spectral function.




To specify the form of $\Phi_{MN}$, we consider the Fierz--Pauli action for a massless, neutral spin-two field,
\begin{equation}
	\label{eq.FP}
	\begin{aligned}
		S_{\Phi}=\frac{1}{2}\int_{\mathcal{M}} & d^{5}x\sqrt{-g}\big[-\nabla_A\Phi_{MN}\nabla^A\Phi^{MN}\\
		&+2\nabla_M\Phi^{MN}\nabla^A\Phi_{AN}+2R_{MNRS}\Phi^{MR}\Phi^{NS}\big].
	\end{aligned}
\end{equation}
We focus on the fields on the fixed AdS$_{5}$ background.

From the equations of motion derived from \eqref{eq.FP}, the solutions are
\begin{equation}
	\label{eq:solution}
	\Phi_{\mu\nu}(r) = \dfrac{\varphi_{\mu\nu}}{r^{2}} + O_{\mu\nu}r^{2}.
\end{equation}
Here, we only consider the leading-order term near the UV boundary, $r\to0$, because its coefficients $\varphi_{\mu\nu}$ are treated as external sources that modify the fermion spectral functions.
From the perspective of experimental results, this setup corresponds to applying external sources to a many-body system.
Accordingly, the leading term provides a direct correspondence between the holographic source parameters $\varphi_{\mu\nu}$ and the experimentally controlled parameters.
Furthermore, this choice renders our interacting spinor model analytically solvable.
For these reasons, we take the ansatz in \eqref{eq2.7} as
\begin{equation}
	\label{eq2.8}
	\Phi_{\mu\nu}(r) = \varphi_{\mu\nu}/r^{2} = \frac{1}{r^{2}}\begin{pmatrix}
		\varphi_{tt} & \varphi_{tx} & \varphi_{ty} & \varphi_{tz}\\
		\varphi_{tx} & \varphi_{xx} & \varphi_{xy} & \varphi_{xz}\\
		\varphi_{ty} & \varphi_{xy} & \varphi_{yy} & \varphi_{yz}\\
		\varphi_{tz} & \varphi_{xz} & \varphi_{yz} & \varphi_{zz}
	\end{pmatrix},
\end{equation}
where $\varphi_{\mu\nu}\in\mathbb{R}$ are constant  parameters that characterize the strength of the symmetric tensor coupling.
Until section~\ref{section4}, we assume $|\varphi_{\mu\nu}| < 1$ for simplicity in examining the effects on spectral densities.
In section~\ref{section3.4p}, we will examine the  case $|\varphi_{ti}| \geq 1$, which gives rise to critically tilted and over-tilted spectral densities.

By \eqref{eq.FP}, the symmetric tensor already possesses traceless and co-divergence-free properties.
These constraints therefore should be considered when we realize and compare our model with experiments. 
Nevertheless, as in \cite{Oh:2020cym,Sukrakarn:2023ncp,Yuk:2022lof,Ghorai:2023wpu,PhysRevD.109.066004,Han:2024rbr}, we classify the role of each individual component in this work for clarity and to provide building blocks for understanding the spectral features.

%

\section{One-flavor spinor}
\label{section3}
This section applies the holographic framework from section~\ref{section2} to derive the analytic retarded Green’s function for a one-flavor spinor under symmetric tensor coupling. 
We further examine how this coupling induces scaling and rotational deformations in the spectral densities, providing a classification of symmetric tensor couplings based on their effects on the spectral density.

\subsection{Dirac equation and source identification}
\label{section3.1}
Throughout this paper, we adopt the gamma matrix representation as
\begin{equation}
	\begin{aligned}
		\label{eq3.1}
		\Gamma^{\underline{t}} &= -i\sigma_{1}\otimes \mathbbm{1}_{2\times2},\,\,\, &\Gamma^{\underline{x}} =& \sigma_{2}\otimes\sigma_{1},\,\,\, &\Gamma^{\underline{y}} &= \sigma_{2}\otimes\sigma_{2},\quad \\
		\Gamma^{\underline{z}} &= \sigma_{2}\otimes\sigma_{3},\,\,\,
		&\Gamma^{\underline{r}} =& \sigma_{3}\otimes\mathbbm{1}_{2\times2},\,\,\,
		&\Gamma^{ab} &= \frac{1}{2}[\Gamma^{a}, \Gamma^{b}].
	\end{aligned}
\end{equation}
The gamma matrices satisfy the Clifford algebra $\{\Gamma^{M}, \Gamma^{N}\} \allowbreak= 2g^{MN} \mathbbm{1}_{4\times4}$ in five dimensions
so that $\Gamma^{MN}$ contains some metric dependence.
We define the metric of the bulk in the inverse radius coordinate as
\begin{equation}
	\label{eq3.2}
	ds^{2} = -\frac{f(r)}{r^{2}}dt^{2} + \frac{dx^{2} + dy^{2} + dz^{2}}{r^{2}} + \frac{dr^{2}}{r^{2}f(r)},
\end{equation}
which is asymptotically $\mathrm{AdS}_{5}$. 
Also, in \eqref{eq2.4}, $\Lambda = -6$ and we set the $\mathrm{AdS}_{5}$ radius to 1 throughout this paper.
From the total action in \eqref{eq2.1}, we can write down the bulk equation of motion for $\psi$ as
\begin{equation}
	\label{eq3.3}
	\begin{gathered}
		(\Gamma^{M}D_{M} - m)\psi + \Phi_{MN}\Gamma^{(M}D^{N)}\psi\\
		 = [(g + \Phi)_{MN}\Gamma^{M}D^{N} - m]\psi = 0,
	\end{gathered}
\end{equation}
We put an ansatz of Dirac field as
\begin{equation}
	\label{eq3.4}
	\psi = (-gg^{rr})^{-1/4}e^{-i\omega t + ik_{x}x + ik_{y}y + ik_{z}z}\zeta(r),
\end{equation}
where $\zeta(r)$ is a four-component spinor field. 
The term $(-gg^{rr})^{-1/4}$ was introduced to remove spin-connection term in Dirac equation, which greatly simplifies 
the system. 
For instance, in pure $\mathrm{AdS}_{5}$ geometry with the symmetric tensor given by \eqref{eq2.8}, the Dirac equation is
\begin{equation}
	\label{eq3.5}
	\begin{gathered}
		\Gamma^{\underline{r}}\partial_{r}\zeta(r) + ir^{2}\Gamma^{\mu}(g + \Phi)_{\mu\nu}k^{\nu}\zeta(r) - \dfrac{\Tr\Phi}{2r}\Gamma^{\underline{r}}\zeta(r) - \dfrac{m}{r}\zeta(r)\\ = 0,\\
		\text{where}\,\, k^{\mu} = (\omega, \boldsymbol{k}) = (\omega, k_{x}, k_{y}, k_{z})\,\,\text{and}\,\,\Tr\Phi = g^{\mu\nu} \Phi_{\mu\nu}(r).
	\end{gathered}
\end{equation}
We will utilize this equation of motion when applying the flow equation and defining the retarded Green’s function in pure $\mathrm{AdS}_{5}$ system.

To define the retarded Green’s function, we first identify the source and condensation. 
For this, we decompose the spinor field in \eqref{eq3.4} as
\begin{equation}
	\label{eq3.6}
	\psi = \begin{pmatrix}
		\psi_{+}\\
		\psi_{-}
	\end{pmatrix}\quad\text{and}\quad\zeta = \begin{pmatrix}
		\zeta_{+}\\
		\zeta_{-}
	\end{pmatrix}.
\end{equation}
Then, by applying the ansatz \eqref{eq3.4}, the boundary action in \eqref{eq2.3} can be rewritten as
\begin{equation}
	\label{eq3.7}
	\begin{aligned}
		S_{\mathrm{bdy}} &= -i\int_{\partial\mathcal{M}}d^{4}x\bar{\zeta}\zeta = -\int_{\partial\mathcal{M}}d^{4}x\zeta^{\dagger}(\sigma_{1}\otimes\sigma_{0})\zeta\\
		  &=  -\int_{\partial\mathcal{M}}d^{4}x\,\zeta^{\dagger}_{+}\zeta_{-} + \mathrm{h.c.}
	\end{aligned}
\end{equation}
By varying the bulk action with respect to $\psi$ and add the variation of the first term in \eqref{eq3.7}, it can be shown that the total action variation can be represented only in terms of $\zeta_{+}$ if the equation of motion is satisfied \cite{Iqbal:2009fd}.
Therefore, we interpret the boundary quantities of $\zeta_{+}$ and $\zeta_{-}$ as the two-component source and condensation, respectively.
For notational clarity, we redefine the bulk quantities $\zeta_{+}$ and $\zeta_{-}$ as $\xi^{(S)}$ and $\xi^{(C)}$, respectively, corresponding to the source and condensation at the boundary:
\begin{equation}
	\label{eq3.8}
	\xi^{(S)} \equiv \zeta_{+}\quad\text{and}\quad\xi^{(C)} \equiv \zeta_{-}.
\end{equation}

To find the retarded Green’s function, we extract the source and condensation terms from their corresponding bulk quantities, $\xi^{(S)}$ and $\xi^{(C)}$.
To achieve this, we examine the boundary behavior of $\xi^{(S)}$ and $\xi^{(C)}$ by solving the Dirac equation in \eqref{eq3.3}.
We can represent the Dirac equation in terms of $\xi^{(S)}$ and $\xi^{(C)}$ in \eqref{eq3.8} as
\begin{gather}
	\label{eq3.9}
	\partial_{r}\xi^{(S)} + \mathbb{M}_{1}\xi^{(S)} + \mathbb{M}_{2}\xi^{(C)} = 0,\\
	\label{eq3.10}
	\partial_{r}\xi^{(C)} + \mathbb{M}_{3}\xi^{(C)} + \mathbb{M}_{4}\xi^{(S)} = 0.
\end{gather}
where $\mathbb{M}_{1}$, $\mathbb{M}_{2}$, $\mathbb{M}_{3}$, and $\mathbb{M}_{4}$ are $2\times2$ matrix-valued functions. 
We consider near-boundary behavior of $\xi^{(S)}$ and $\xi^{(C)}$ for pure $\mathrm{AdS}_{5}$ case by analytically solving \eqref{eq3.9} and \eqref{eq3.10}.
For spinors with $\abs{m} < 1/2$, the leading $r$-dependent terms of the spinors $\xi^{(S)}$ and $\xi^{(C)}$ near the boundary are extracted as
\begin{equation}
	\begin{aligned}
		\label{eq3.11}
		\xi^{(S)} \approx r^{m + \Tr\varphi/2}\mathcal{J}\quad&\text{and}\quad\xi^{(C)} \approx r^{-m + \Tr\varphi/2}\mathcal{C},\\
		\text{where}\quad\Tr\varphi = \eta^{\mu\nu}\varphi_{\mu\nu}\quad&\text{and}\quad\eta_{\mu\nu} = \mathrm{diag}(-1, 1, 1, 1).
	\end{aligned}
\end{equation}
Here, $\varphi_{\mu\nu}$ is defined in \eqref{eq2.8}, $\mathcal{J}$ and $\mathcal{C}$ are two-component spinors, representing the source and condensation, respectively.
We will use these source and condensation terms to determine the retarded Green’s function for one-flavor spinors in the following section.

\subsection{Definition of retarded Green's function}
\label{section3.3}
When we define the retarded Green's function from our chosen source and condensation, we adopt the formalism in \cite{Yuk:2022lof} to express the retarded Green's function and  solve it analytically instead of  solving the Dirac equation.

Because $\xi^{(S)}$ and $\xi^{(C)}$ are two-component spinors, each has two independent solutions. 
The general solution can therefore be expressed as a linear combination of these two solutions with constant coefficients.
It is convenient to represent $\xi^{(S)}$ and $\xi^{(C)}$ in matrix form, separating the solution basis from their coefficients.
For example, if we denote the two basis solutions for $\xi^{(S)}$ as $(\xi^{(S, 1)}_{1}, \xi^{(S, 1)}_{2})^T$ and $(\xi^{(S, 2)}_{1}, \xi^{(S, 2)}_{2})^T$, with the corresponding coefficients $c_{1}$ and $c_{2}$, then the spinor $\xi^{(S)}$ can be expressed as
\begin{equation}
	\begin{adjustbox}{max width=\columnwidth}
	$\begin{gathered}
		\label{eq3.12}
		\xi^{(S)} = c_{1}\begin{pmatrix}
			\xi^{(S, 1)}_{1}\\
			\xi^{(S, 1)}_{2}
		\end{pmatrix} + c_{2}\begin{pmatrix}
			\xi^{(S, 2)}_{1}\\
			\xi^{(S, 2)}_{2}
		\end{pmatrix} = \begin{pmatrix}
			\xi^{(S, 1)}_{1} & \xi^{(S, 2)}_{1}\\
			\xi^{(S, 1)}_{2} & \xi^{(S, 2)}_{2}
		\end{pmatrix}\begin{pmatrix}
			c_{1}\\
			c_{2}
		\end{pmatrix} = \mathbb{S}(r){\bf c},\\
		\text{where}\quad\mathbb{S}(r) = \begin{pmatrix}
			\xi^{(S, 1)}_{1} & \xi^{(S, 2)}_{1}\\
			\xi^{(S, 1)}_{2} & \xi^{(S, 2)}_{2}
		\end{pmatrix}\quad\text{and}\quad{\bf c} = \begin{pmatrix}
			c_{1}\\
			c_{2}
		\end{pmatrix}.
	\end{gathered}$
\end{adjustbox}
\end{equation}
Likewise, we can represent the $\xi^{(C)}$ in a similar way. 
Meanwhile, due to the Dirac equations from \eqref{eq3.9} to \eqref{eq3.10}, $\xi^{(C)}$ can be expressed in terms of $\xi^{(S)}$; that is, $\xi^{(C)}$ shares the same coefficient   ${\bf c}$ as $\xi^{(S)}$, such that
\begin{equation}
	\label{eq3.13}
	\xi^{(C)} = \mathbb{C}(r){\bf c},
\end{equation}
where $\mathbb{C}(r)$ is a $2\times2$ matrix-valued function. 
From the boundary behavior of $\xi^{(S)}$ and $\xi^{(C)}$ in \eqref{eq3.11}, we see that the leading terms $r^{\pm m + \Tr\varphi/2}$ of $\xi^{(S)}$ and $\xi^{(C)}$ at the boundary originate from those of $\mathbb{S}(r)$ and $\mathbb{C}(r)$ in \eqref{eq3.12} and \eqref{eq3.13}, respectively, because ${\bf c}$ is a constant vector.
Therefore, the boundary behavior of $\mathbb{S}(r)$ and $\mathbb{C}(r)$ is given by 
\begin{equation}
	\label{eq3.14}
	\mathbb{S}(r) \approx r^{m + \Tr\varphi/2}\mathbb{S}_{0}\quad\text{and}\quad\mathbb{C}(r) \approx r^{-m + \Tr\varphi/2}\mathbb{C}_{0},
\end{equation}
with constant matrices $\mathbb{S}_{0}$ and $\mathbb{C}_{0}$.
Then, the spinors $\xi^{(S)}$ and $\xi^{(C)}$ are given  by 
\begin{equation}
	\label{eq3.15}
	\xi^{(S)} \approx r^{m + \Tr\varphi/2}\mathbb{S}_{0}{\bf c}\quad\text{and}\quad\xi^{(C)} \approx r^{-m + \Tr\varphi/2}\mathbb{C}_{0}{\bf c}. 
\end{equation}
Comparing \eqref{eq3.11} with \eqref{eq3.15}, the source and condensation can be rewritten as
\begin{equation}
	\label{eq3.16}
	\mathcal{J} = \mathbb{S}_{0}{\bf c}\quad\text{and}\quad\mathcal{C} = \mathbb{C}_{0}{\bf c} \quad \hbox{ so that }  \mathcal{C} = \mathbb{C}_{0}\mathbb{S}^{-1}_{0}\mathcal{J}. 
\end{equation}

We now can get the retarded Green's function in terms of $\mathbb{S}_{0}$ and $\mathbb{C}_{0}$ from the total action. 
Meanwhile, since the bulk action does not contribute to the total action due to the equation of motion, the Green’s function is determined solely from the boundary action in \eqref{eq3.7}.
To achieve this, we rewrite the boundary action, which is denoted as the effective action $S_{\mathrm{eff}}$, near the boundary at $r = \epsilon$, in terms of $\mathcal{J}$ and $\mathcal{C}$ as
\begin{equation}
	\label{eq3.17}
	\begin{aligned}
		S_{\mathrm{eff}} &= -\int_{\partial\mathcal{M}}d^{4}x\,{\xi^{(S)}}^{\dagger}\xi^{(C)} + \mathrm{h.c.}\\
		&= -\int_{\partial\mathcal{M}}d^{4}x\,\epsilon^{\Tr\varphi}\mathcal{J}^{\dagger}\mathcal{C} + \mathrm{h.c.}
	\end{aligned}
\end{equation}
From the   equation  \eqref{eq3.16},   the effective action can be rewritten  in terms of the source only,   
\begin{equation}
	\label{eq3.18}
	S_{\mathrm{eff}} = -\int_{\partial\mathcal{M}}d^{4}x\,\epsilon^{\Tr\varphi}\mathcal{J}^{\dagger}(\mathbb{C}_{0}\mathbb{S}^{-1}_{0})\mathcal{J} + \mathrm{h.c.}
\end{equation}
According to the linear response theory, the effective action can be represented as
\begin{equation}
	\label{eq3.19}
	S_{\mathrm{eff}} = -\int_{\partial\mathcal{M}}d^{4}x\,\epsilon^{\Tr\varphi}\mathcal{J}^{\dagger}G_{R}\mathcal{J} + \mathrm{h.c.}
\end{equation}
where $G_{R}$ is the retarded Green's function for the source $\mathcal{J}$. By comparing \eqref{eq3.18} and \eqref{eq3.19}, we identify the  matrix-valued retarded Green's function as 
\begin{equation}
	\label{eq3.20}
	G_{R} = \mathbb{C}_{0}\mathbb{S}^{-1}_{0}.
\end{equation}
On the other hand,   the flow equation provides the bulk quantity $\mathbb{G}(r)$ defined by 
\begin{equation}
	\label{eq3.21}
	\mathbb{G}(r) \equiv \mathbb{C}(r)\mathbb{S}^{-1}(r),
\end{equation}
whose boundary behavior will give the retarded Green’s function $G_{R}$. 
Thus, to determine $G_{R}$ from $\mathbb{G}(r)$, we should look for  the boundary behavior of $\mathbb{G}(r)$.
From \eqref{eq3.14} and \eqref{eq3.20},   
\begin{equation}
	\label{eq3.22}
	\mathbb{G}(r) = \mathbb{C}(r)\mathbb{S}^{-1}(r) \approx r^{-2m}\mathbb{C}_{0}\mathbb{S}^{-1}_{0} = r^{-2m}G_{R}.
\end{equation}
Then, we can find the retarded Green's function from the bulk quantity $\mathbb{G}(r)$ as
\begin{equation}
	\label{eq3.23}
	G_{R} = \lim_{r\to0}r^{2m}\mathbb{G}(r).
\end{equation}
Therefore, once we  know     $\mathbb{G}(r)$, we   can calculate $G_{R}$. This is the calculational basis of our work. 

{Notice that from  \eqref{eq3.23}, we see that $\Tr\varphi/2$   in the boundary behavior of $\xi^{(S)}$ and $\xi^{(C)}$ does not affect to the definition of retarded Green's function. Also, for   spinors without the bulk mass, the retarded Green's function is defined by the leading term of $\mathbb{G}(r)$ at the boundary.}

\subsection{Spectral densities in pure $\mathrm{AdS}_{5}$}
\label{section3.4}
If we find the equation satisfied by $\mathbb{G}(r)$ satisfying certain boundary condition at the IR, we can derive the retarded Green’s function in \eqref{eq3.20} by solving
it. It turns out that such equation is  a first-order differential equation called flow equation \cite{PhysRevD.79.025023,Yuk:2022lof}. 
We   adopt the  formalism in \cite{Yuk:2022lof}  to  take the advantage that we can get the analytic Green's function  directly without explicitly solving for the spinors $\xi^{(S)}$ and $\xi^{(C)}$.

The flow equation can be set up using   \eqref{eq3.9} and \eqref{eq3.10}. For the explicit derivation, we refer the reader  to  the Appendix~\ref{A1} at the end of this paper.
The   flow equation for the bulk quantity $\mathbb{G}(r)$ is given by
\begin{equation}
	\label{eq3.24}
	\begin{gathered}
		\partial_{r}\mathbb{G}(r) + \mathbb{G}(r)\bar{\mathbb{M}}_{2}\mathbb{G}(r) + \mathbb{G}(r)\bar{\mathbb{M}}_{1} + \mathbb{M}_{3}\mathbb{G}(r) + \mathbb{M}_{4} = 0,\\
		\text{where}\quad\bar{\mathbb{M}}_{1} = -\mathbb{M}_{1}\quad\text{and}\quad \bar{\mathbb{M}}_{2} = -\mathbb{M}_{2}.
	\end{gathered}
\end{equation}
This is a matrix version of Riccati equation introduced in \cite{PhysRevD.79.025023}.
To solve this flow equation analytically, we restrict our analysis to the probe level analysis where we set the metric as the  pure $\mathrm{AdS}_{5}$
\begin{equation}
	\label{eq3.25}
	ds^2= \dfrac{1}{r^{2}}\left(-dt^{2} + dx^{2} + dy^{2} + dz^{2} + dr^{2}\right).
\end{equation}
In this case, the four matrices in \eqref{eq3.24} are given by
\begin{equation}
	\label{eq3.26}
	\begin{gathered}
		\bar{\mathbb{M}}_{1} = \left(\dfrac{m}{r} + \Tr\Phi/2\right)\mathbbm{1}_{2\times2},\quad \bar{\mathbb{M}}_{2} = -r^{2}\sigma^{\mu}(g + \Phi)_{\mu\nu}k^{\nu},\\
		\mathbb{M}_{3} = \left(\dfrac{m}{r} - \Tr\Phi/2\right)\mathbbm{1}_{2\times2},\quad \mathbb{M}_{4} = -r^{2}\bar{\sigma}^{\mu}(g + \Phi)_{\mu\nu}k^{\nu},\\
		\text{with}\quad\sigma^{\mu} = (\sigma_{0}, \sigma^{i})\quad\text{and}\quad\bar{\sigma} = (\sigma_{0}, -\sigma^{i}). 
	\end{gathered}
\end{equation}
Notice  that the symmetric tensor couples with     four-momentum  with help of the metric.
When we substitute \eqref{eq3.26} into \eqref{eq3.24}, the flow equation becomes
\begin{equation}
	\label{eq3.27}
	\begin{aligned}
		\partial_{r}\mathbb{G}(r) &- r^{2}\mathbb{G}(r)\sigma^{\mu}\mathbb{G}(r)(g + \Phi)_{\mu\nu}k^{\nu} + \frac{2m}{r}\mathbb{G}(r)\\
		&- r^{2}\bar{\sigma}^{\mu}(g + \Phi)_{\mu\nu}k^{\nu} = 0.
	\end{aligned}
\end{equation}
To solve this flow equation, we impose the IR boundary condition for $\mathbb{G}(r)$, which is given by~\cite{Yuk:2022lof}
\begin{equation} 
	\label{eq3.28}
	\mathbb{G}(r\to\infty) = i\mathbbm{1}_{2\times2}.
\end{equation}
This boundary condition together with the flow equation determines the  $\mathbb{G}(r)$ 
uniquely. 

Below, we  present  an analytical solution  whose detailed derivation is  included in Appendix~\ref{A1}: 
\begin{equation}
	\label{eq3.29}
	\begin{gathered}
		\mathbb{G}(r) = -\dfrac{K_{m + \frac{1}{2}}\left(|(\eta + \varphi)_{\mu\nu}k^{\nu}|r\right)}{K_{m - \frac{1}{2}}\left(|(\eta + \varphi)_{\mu\nu}k^{\nu}|r\right)}\dfrac{\bar{\sigma}^{\mu}(\eta + \varphi)_{\mu\nu}k^{\nu}}{|(\eta + \varphi)_{\mu\nu}k^{\nu}|}\\
		\text{where}\quad|(\eta + \varphi)_{\mu\nu}k^{\nu}| = [\eta^{\mu\nu}(\eta + \varphi)_{\mu\rho}k^{\rho}(\eta + \varphi)_{\nu\sigma}k^{\sigma}]^{1/2}.
	\end{gathered}
\end{equation}
Here we select the branch as $-\pi/2 < \arg|(\eta + \varphi)_{\mu\nu}k^{\nu}| \leq \pi/2$.
From this $\mathbb{G}(r)$ the retarded Green’s function can be derived using \eqref{eq3.23}.
In the massless case ($m = 0$), the retarded Green’s function simplifies to
\begin{equation}
	\label{eq3.30}
	G_{R}(\omega, \boldsymbol{k}) = -\frac{\bar{\sigma}^{\mu}(\eta + \varphi)_{\mu\nu}k^{\nu}}{|(\eta + \varphi)_{\mu\nu}k^{\nu}|}.
\end{equation}
The trace of this retarded Green's function is 
\begin{equation}
	\label{eq3.31}
	\Tr G_{R}(\omega, \boldsymbol{k}) = \frac{-2(\eta + \varphi)_{t\mu}k^{\mu}}{\left|(\eta + \varphi)_{\nu\rho}k^{\rho}\right|}.
\end{equation}
To understand the effects of symmetric tensor couplings on the   Green’s function, it is useful to compare it with that  of free fermions.
When we let $\varphi_{\mu\nu} = 0$ in \eqref{eq3.30}, we recover the Green's 
function for massless free spinors\cite{Iqbal:2009fd}: 
\begin{equation}
	\label{eq3.32}
	G_{R, \mathrm{free}}(\omega, \boldsymbol{k}) = \frac{1}{\sqrt{|\boldsymbol{k}|^{2} - \omega^{2}}}\begin{pmatrix}
		k_{z} + \omega & k_{x} - ik_{y}\\
		k_{x} + ik_{y} & -k_{z} + \omega
	\end{pmatrix},
\end{equation}
whose  trace is given by
\begin{equation}
	\label{eq3.33}
	\Tr G_{R, \mathrm{free}} = \frac{2\omega}{\sqrt{|\boldsymbol{k}|^{2} - \omega^{2}}}.
\end{equation}

\begin{figure*}[tb]
	\centering
	\includegraphics[width =0.8\linewidth]{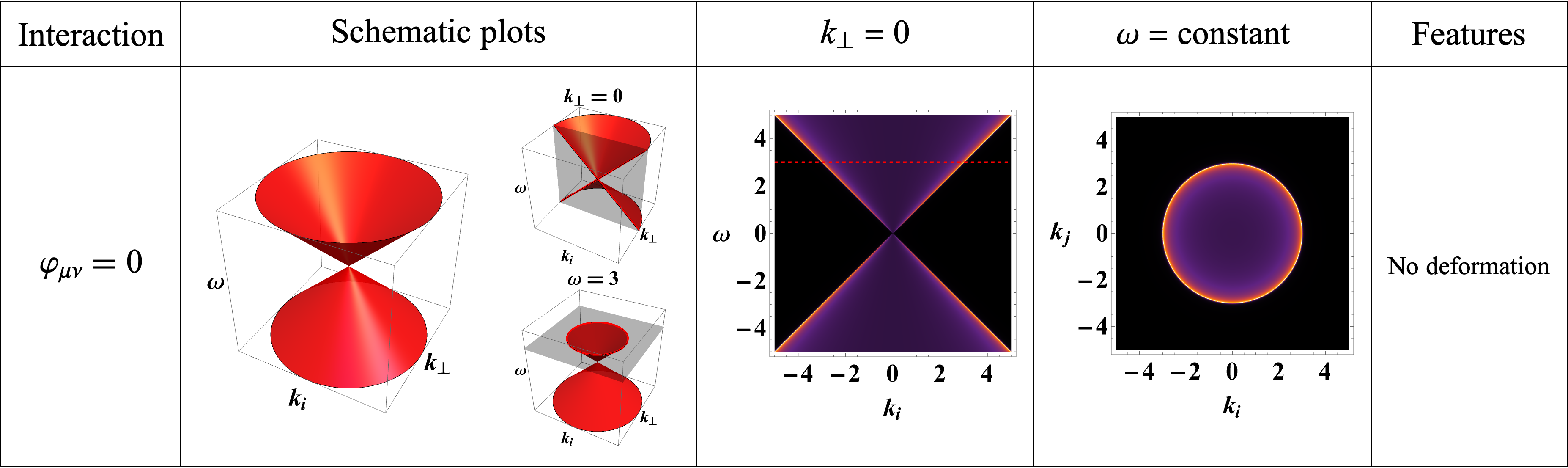}
	\caption{Schematic plots of the spectral density (SD) and its cross-sectional slices for free and massless fermions.}
	\label{figure2}
\end{figure*}

The spectral density is defined as the imaginary part of traced retarded Green's 
function: 
\begin{equation}
	\label{eq3.34}
	A(\omega, \boldsymbol{k}) = \mathrm{Im}\left(\Tr G_{R}(\omega, \boldsymbol{k})\right).
\end{equation}
We  get the  spectral density  by taking  the imaginary part of \eqref{eq3.33} {\it after taking} $\omega\to\omega + i\epsilon$ ($\epsilon > 0$). For the result, see the figure~\ref{figure2}. One should compare all    other cases  of interacting spinors with this.  

If we turn on each component of the symmetric tensor coupling in \eqref{eq3.30}, we  obtain the corresponding retarded Green’s function and the associated spectral density using  \eqref{eq3.31} and \eqref{eq3.34}. 
We classify the interaction type into $\Phi_{tt}$, $\Phi_{ti}$, $\Phi_{ii}$, and $\Phi_{ij}$ $(i \neq j)$ with $i, j = {x, y, z}$. 
Let  $k_{i}$ and $k_{j}$  be   three-momentum components associated with the indices of  $\Phi_{ti}$, $\Phi_{ii}$, and $\Phi_{ij}$, and  $k_{\perp}$ be  the remaining three-momentum components.
That is,  for  $\Phi_{tx}$, $k_{i} = k_{x}$ and $k_{\perp} = \{k_{y}, k_{z}\}$, whereas for   $\Phi_{xy}$, ${k_{i}, k_{j}} = {k_{x}, k_{y}}$ and $k_{\perp} = k_{z}$.

\begin{figure*}[tb]
	\centering
	\includegraphics[width =0.8\linewidth]{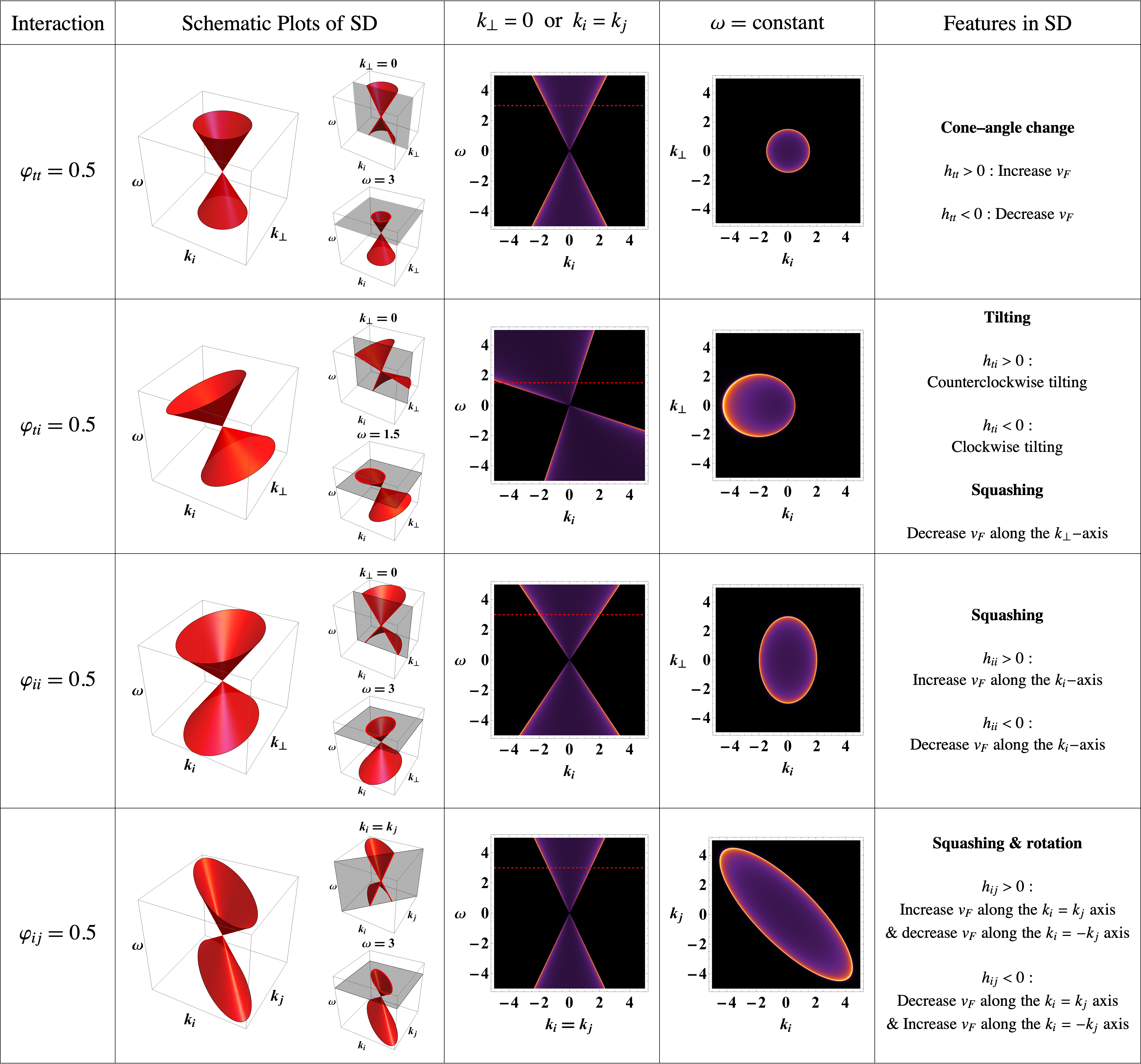}
	\caption{Classification of the spectral density (SD) for massless one-flavor fermions under each type of symmetric tensor coupling. Schematic plots of the SD with $\varphi_{\mu\nu} = 0.5$ and their cross-sectional slices (gray planes and red dotted lines) are shown.}
	\label{figure3}
\end{figure*}

It turns out that for all cases, the Green's functions of interacting case can be obtained from that of the free fermion by a simple transformation of $(\omega,k_i)$. We tabulated the result in the table~\ref{table2}. 
\begin{table*}
	\centering\footnotesize
	\resizebox{0.8\linewidth}{!}{
		\begin{tabular}[c]{| c | c | c|}
			\hline
			\textbf{Int.} & \textbf{Trace of analytic retarded Green's function} & \textbf{Transformation}\\
			\hline
			$\Phi_{tt}$ & \begin{minipage}{0.61\textwidth}\vspace{5pt}
				\begin{equation}
					\label{eq3.35}
					\Tr G_{R} = \dfrac{2\omega}{\sqrt{\dfrac{|\boldsymbol{k}|^{2}}{(1 - \varphi_{tt})^{2}} - \omega^{2}}}
				\end{equation}\vspace{0pt}
			\end{minipage} & $
			\begin{gathered}
				\boldsymbol{k}\to\dfrac{\boldsymbol{k}}{1 - \varphi_{tt}}\\
				\text{or}\\
				\omega\to(1 - \varphi_{tt})\omega
			\end{gathered}
			$\\
			\hline
			$\Phi_{ti}$ & \begin{minipage}{0.61\textwidth}\vspace{5pt}
				\begin{equation}
					\label{eq3.36}
					\Tr G_{R} = \dfrac{2(\omega - \varphi_{ti} k_{i})}{\sqrt{
							\begin{pmatrix} \omega & k_{i} \end{pmatrix}
							\begin{pmatrix}
								-1 + \varphi_{ti}^{2} & 2 \varphi_{ti} \\
								2 \varphi_{ti} & 1 - \varphi_{ti}^{2}
							\end{pmatrix}
							\begin{pmatrix} \omega \\ k_{i} \end{pmatrix}
							+ |\boldsymbol{k}_{\perp}|^{2}}}
				\end{equation}\vspace{0pt}
			\end{minipage} & $\begin{pmatrix}
				\omega\\
				k_{i}
			\end{pmatrix}\to\begin{pmatrix}
				1 & -\varphi_{ti}\\
				\varphi_{ti} & 1
			\end{pmatrix}\begin{pmatrix}
				\omega\\
				k_{i}
			\end{pmatrix}$\\
			\hline
			$\Phi_{ii}$ & \begin{minipage}{0.61\textwidth}\vspace{5pt}
				\begin{equation}
					\label{eq3.37}
					\Tr G_{R} = \dfrac{2\omega}{\sqrt{(1 + \varphi_{ii})^{2} k_{i}^{2} + |\boldsymbol{k}_{\perp}|^{2} - \omega^{2}}}
				\end{equation}\vspace{0pt}
			\end{minipage} & $k_{i}\to(1 + \varphi_{ii})k_{i}$\\
			\hline
			$\Phi_{ij}$ & \begin{minipage}{0.61\textwidth}
				\begin{equation}
					\label{eq3.38}
					\Tr G_{R} = \dfrac{2\omega}{\sqrt{
							\begin{pmatrix} k_{i} & k_{j} \end{pmatrix}
							\begin{pmatrix}
								1 + \varphi_{ij}^{2} & 2 \varphi_{ij} \\
								2 \varphi_{ij} & 1 + \varphi_{ij}^{2}
							\end{pmatrix}
							\begin{pmatrix} k_{i} \\ k_{j} \end{pmatrix}
							+ k_{\perp}^{2} - \omega^{2}}}
				\end{equation}\vspace{0pt}
			\end{minipage} & $
			\begin{gathered}
				\begin{pmatrix}
					k_{i}\\
					k_{j}
				\end{pmatrix}\to R_{\frac{\pi}{4}}^{T}S_{\varphi}R_{\frac{\pi}{4}}\begin{pmatrix}
					k_{i}\\
					k_{j}
				\end{pmatrix}\\
				\begin{cases}
					R_{\frac{\pi}{4}} : \text{rotation by $\pi/4$}\\
					S_{\varphi} = \mathrm{diag}(1 - \varphi_{ij}, 1 + \varphi_{ij})
				\end{cases}
			\end{gathered}$\\
			\hline
		\end{tabular}
	}
	\caption{Traces of the analytic retarded Green’s function and the corresponding momentum transformations for one-flavor spinors under each interaction type (Int.) of symmetric tensor couplings.}
	\label{table2}
\end{table*}
Note that all $\Tr G_{R}$ have a branch-cut singularity. 
By taking the imaginary part of $\Tr G_{R}$  with    $\omega\to\omega + i\epsilon$ ($\epsilon > 0$), we classify four types of SD in the figure~\ref{figure3} based on their behavior. 

In figure 3, the Fermi velocity, $v_{F}$, is defined from the spectral densities as the slope of the SD’s shell in momentum space, 
\begin{equation}
	v_{F} = \nabla_{\boldsymbol{k}}E(\boldsymbol{k}),
\end{equation}
where $E(\boldsymbol{k})$ represents the dispersion relation derived from the pole structure of $\Tr G_R$.

As depicted in the table~\ref{table2} and figure~\ref{figure3}, the spectral features associated with each symmetric tensor   can be   classified into  a few categories: cone-angle change, tilting, squashing  and rotation: 
\begin{itemize}
	\item 
	The $\Phi_{tt}$ component modifies the Fermi velocity uniformly, increasing it when $\Phi_{tt} > 0$ and decreasing it when $\Phi_{tt} < 0$, leading to cone-angle change on the spectral density.
	\item 	
	The $\Phi_{ti}$ components induce tilting of spectral densities in the $\omega\text{-}k_{i}$ plane, causing a counterclockwise tilting for $\Phi_{ti} > 0$ and a clockwise tilting for $\Phi_{ti} < 0$, while preserving the $\pi/2$ cone angle in the same plane.
	Additionally, these components induce squashing effects on the spectral densities along each $k_{\perp}$-axis, resulting in a reduction of the Fermi velocity along that axis. 
	This, in turn, increases the cone angle in the corresponding direction, independent of the sign of $\varphi_{ti}$.
	Overall, the $\Phi_{ti}$ component induces asymmetry in the spectral density, including both tilting and squashing effects.
	\item 	
	The $\Phi_{ii}$ components either increase ($\Phi_{ii} > 0$) or decrease ($\Phi_{ii} < 0$) the Fermi velocity along the $k_{i}$-axis, resulting in squashing effects on the spectral densities in the $k_{i}$-direction.
	\item 	
	The $\Phi_{ij}$ components increase the Fermi velocity along the $k_{i} = k_{j}$ axis while decreasing it along the orthogonal axis ($k_{i} = -k_{j}$) when $\Phi_{ij} > 0$, and the effect is reversed for $\Phi_{ij} < 0$.
	As shown in the table~\ref{table2}, this coupling can be interpreted as inducing a squashing ($S_{\varphi}$ in the table~\ref{table2}) and rotation $\pi/4$ ($R_{\frac{\pi}{4}}$ in the table~\ref{table2}) effects on the spectral densities.
\end{itemize}
In summary, symmetric tensor couplings induce cone-angle change, squashing, and tilting effects on the spectral densities.

\subsection{\label{section3.4p}Over-tilted spectral density}
\begin{figure*}[tb]
	\centering
	\includegraphics[width =0.8\linewidth]{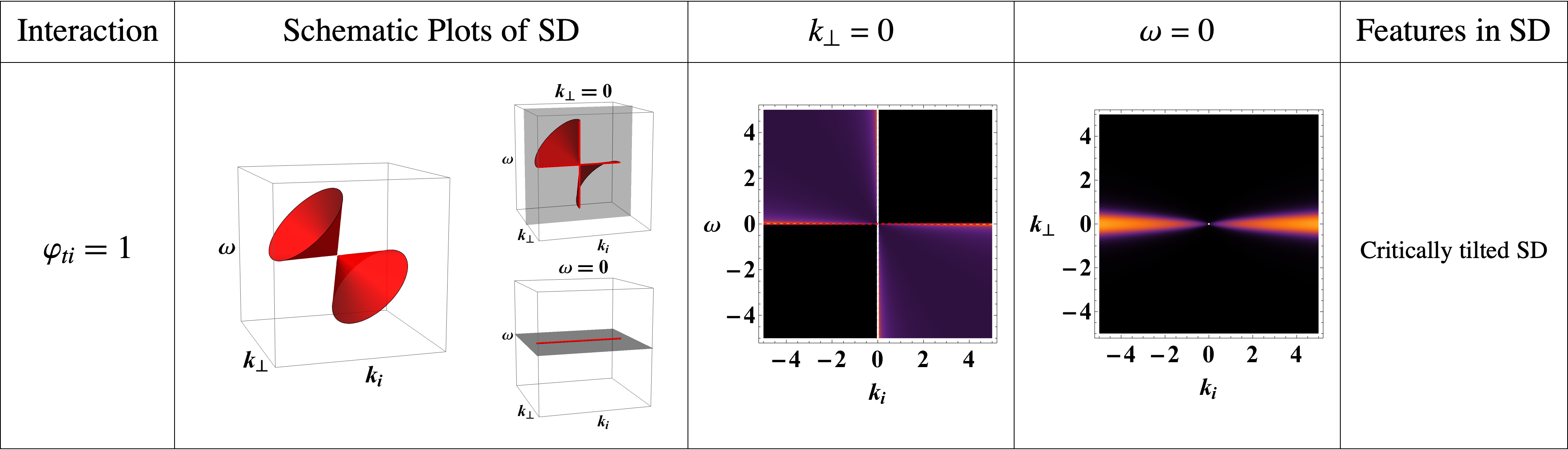}
	\caption{Schematic plots of the critically-tilted spectral density (SD) and its cross-sectional slices (gray planes and red dotted line) for $\varphi_{ti} = 1$.}
	\label{figure6}
\end{figure*}

There are some remarks on $\Phi_{ti}$ coupling.
Within the range $0 \leq |\varphi_{ti}| \leq 1$, the tilted spectral density is well-defined, as it remains positive across the entire momentum space.
Also, at $\varphi_{ti} = 1$, the spectral density exhibits critical tilting, as shown in the figure~\ref{figure6}.

\begin{figure*}[tb]
	\centering
	\includegraphics[width =0.8\linewidth]{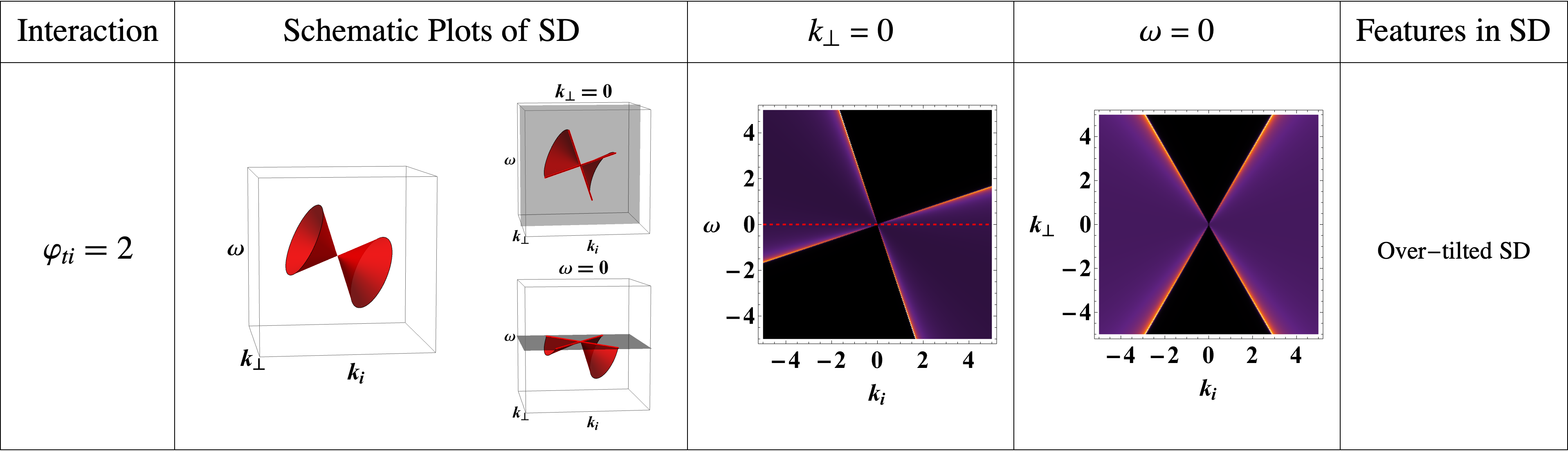}
	\caption{Schematic plots of the over-tilted spectral density (SD) and its cross-sectional slices (gray planes and red dotted line) for $\varphi_{ti} = 2$.}
	\label{figure8}
\end{figure*}

Meanwhile, one technical issue arises for the over-tilted cone with $\abs{\varphi_{ti}} > 1$: taking $\omega \to \omega + i\epsilon$ in \eqref{eq3.36} leads to a negative spectral density due to the incorrect $i\epsilon$ prescription.
To impose the correct prescription for the over-tilted SD, we consider the $i\epsilon$ prescription in a transformed momentum space.
That means, referring to table~\ref{table2}, we consider the following transformation:
\begin{equation}
	\label{eq3.40p}
	\begin{pmatrix}
		\tilde{\omega}\\
		\tilde{k}_{i}
	\end{pmatrix} = \begin{pmatrix}
		1 & -\varphi_{ti}\\
		\varphi_{ti} & 1
	\end{pmatrix}\begin{pmatrix}
		\omega\\
		k_{i}
	\end{pmatrix}.
\end{equation}
This reformulates the Dirac equation for $\Phi_{ti}$ into an equivalent form of that for free spinor cases ($\varphi_{\mu\nu} = 0$), resulting in a traced Green’s function expressed as
\begin{equation}
	\Tr G_{R} = \frac{2\tilde{\omega}}{\sqrt{\tilde{k}_{i}^{2} + \abs{\boldsymbol{k}_{\perp}}^{2} - \tilde{\omega}^{2}}},
\end{equation}
which matches the standard form of the spectral density for free spinors in the transformed momentum space $(\tilde{\omega}, \tilde{k}_{i})$.
We then apply $\tilde{\omega} \to \tilde{\omega} + i\epsilon$ to this traced Green’s function, followed by an inverse transformation of \eqref{eq3.40p} in terms of $\omega$ and $k_{i}$.
The spectral density in this case is always positive even for $\abs{\varphi_{ti}} > 1$.
We draw  this  for $\varphi_{ti} = 2$ in the figure~\ref{figure8}.

We examine the causality of the over-tilted Dirac cones under the above generalized prescription.
For simple analysis, we set $\boldsymbol{k}_{\perp} = 0$.
Under the transformation in \eqref{eq3.40p}, the associated transformation in $(t, x_{i})$ is
\begin{equation}
	\label{eq55p}
	\tilde{t} = \dfrac{t + \varphi_{ti}x_{i}}{1 + \varphi_{ti}^{2}},
	\qquad
	\tilde{x}_{i} = \dfrac{x_{i} - \varphi_{ti}t}{1 + \varphi_{ti}^{2}}.
\end{equation}
Therefore, we examine the causality of the Fourier-transformed Green's function
\begin{equation}
	\begin{aligned}
		G_{R}(t, x_{i}) &= \int\dfrac{d^{2}k}{(2\pi)^{2}}\, e^{-i\omega t + ik_{i}x_{i}}G_{R}(\omega, k_{i})\\
		&= \dfrac{1}{1+\varphi_{ti}^{2}}\int\dfrac{d\tilde{\omega}d\tilde{k}_{i}}{(2\pi)^{2}}\, e^{-i\tilde{\omega}\tilde{t} + i\tilde{k}_{i}\tilde{x}_{i}}G_{R, {\rm free}}(\tilde{\omega}, \tilde{k}_{i}),
	\end{aligned}
\end{equation}
where we have used the transformations \eqref{eq3.40p} and \eqref{eq55p}.
We then apply the above generalized retarded/advanced prescription $\tilde{\omega}\to\tilde{\omega} \pm i\epsilon$.
In the present analysis, we only consider the retarded one.
Because the singularities lie at $\tilde{\omega} = \pm\tilde{k}_{i} - i\epsilon$ and $G_{R, {\rm free}}(\tilde{\omega}, \tilde{k}_{i})$ is analytic for $\Im\tilde{\omega} > 0$, the Green's function vanishes for $\tilde{t} < 0$.
This shows that the Green's function possesses the correct causality with respect to $\tilde{t}$ and identifies $\tilde{t}$ as the physical time in the over-tilted regime.



\begin{figure*}[tb]
	\centering
	\begin{subfigure}[b]{0.8\linewidth}
		\centering
		\includegraphics[width =0.6\textwidth]{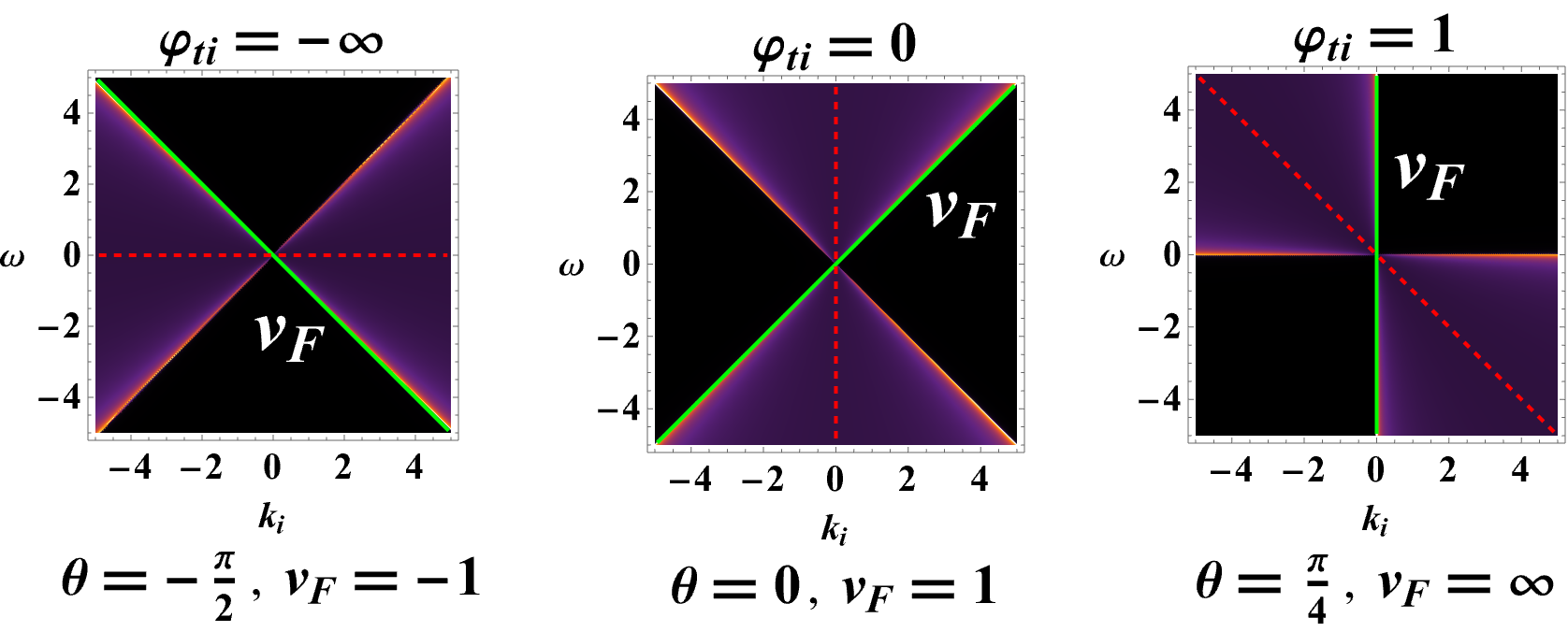}
		\caption{\centering{Definition of the counterclockwise tilting angle $\theta$ (red, dashed) and the Fermi velocity (green, solid) within the region where the interaction strength satisfies $\varphi_{ti} \leq 1$.}}
		\label{figure9a}
	\end{subfigure}\\
	\vspace{5pt}
	\begin{subfigure}[b]{0.28\textwidth}
		\centering
		\includegraphics[width =1.0\textwidth]{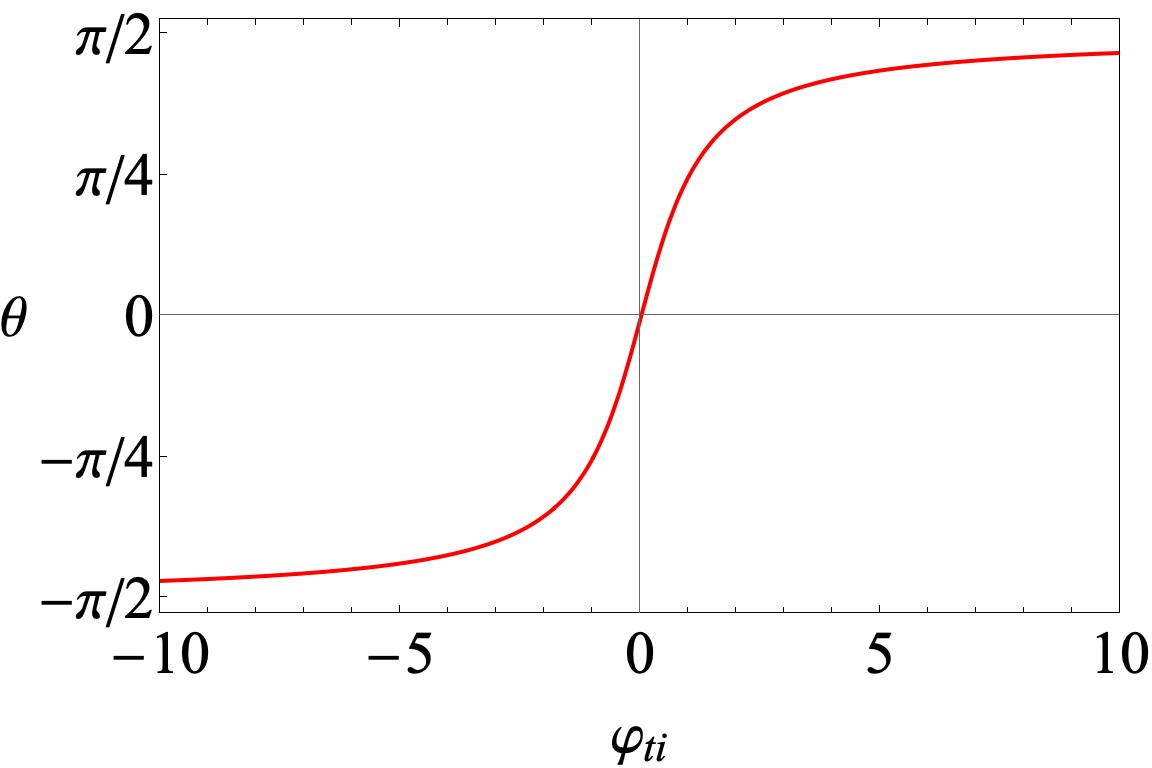}
		\caption{\centering{SD's rotated angle $\theta$ as a function of $\varphi_{ti}$.}}
		\label{figure9b}
	\end{subfigure}
	\begin{subfigure}[b]{0.28\textwidth}
		\centering
		\includegraphics[width=1.0\textwidth]{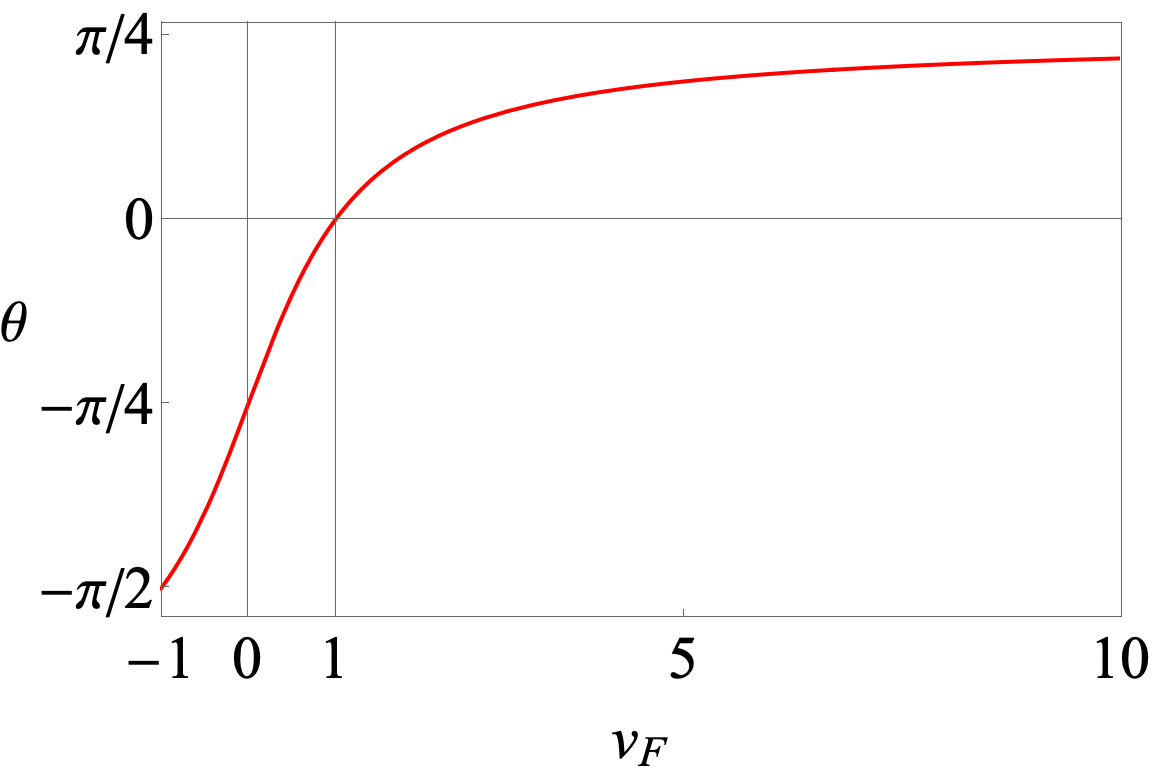}
		\caption{\centering{SD's rotated angle $\theta$ as a function of $v_{F}$.}}
		\label{figure9c}
	\end{subfigure}
	\begin{subfigure}[b]{0.28\textwidth}
		\centering
		\includegraphics[width=1.0\textwidth]{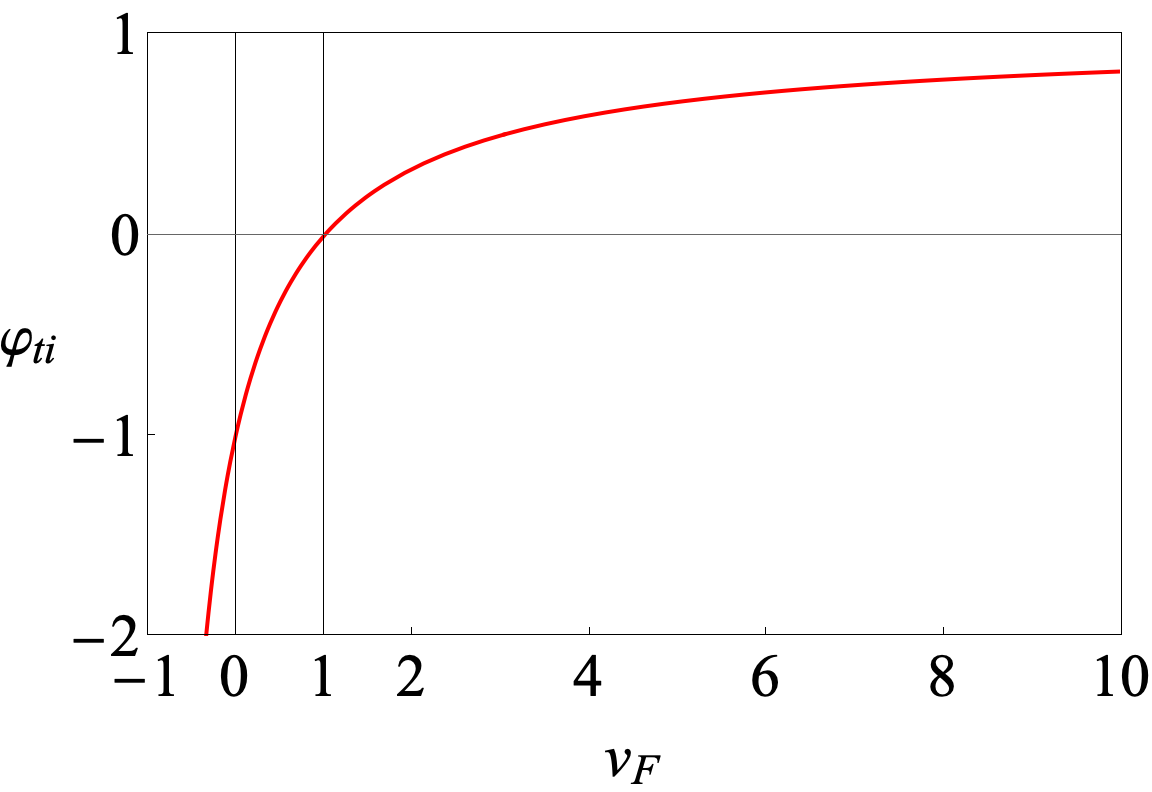}
		\caption{\centering{SD's corresponding $\varphi_{ti}$ as a function of $v_{F}$.}}
		\label{figure9d}
	\end{subfigure}
	\caption{Relations between the Fermi velocity $v_{F}$, the tilting angle $\theta$ of the spectral density, and the coupling strength $\varphi_{ti}$.}
	\label{figure9}
\end{figure*}

Now, we  measure the tilting  angle of the  spectral density relative to the $\omega$-axis for a given $\varphi_{ti}$ value. The tilting  angle is taken to be positive for a positive interaction.  
See the figure~\ref{figure9a}. 
By solving the dispersion relation in \eqref{eq3.36}, we derive the tilting angle $\theta$ of the spectral density as a function of $\varphi_{ti}$:
\begin{equation}
	\label{eq5.8}
	\theta = \begin{cases}
		-\dfrac{\pi}{4} + \arctan\left(\dfrac{1 + \varphi_{ti}}{1 - \varphi_{ti}}\right), & \text{for}\,\,\, \varphi_{ti} \leq 1,\\[15pt]
		\dfrac{3\pi}{4} + \arctan\left(\dfrac{1 + \varphi_{ti}}{1 - \varphi_{ti}}\right), & \text{for}\,\,\, \varphi_{ti} > 1.\\
	\end{cases}
\end{equation}
We present this relation in the figure~\ref{figure9b}.
From this figure, we see that the tilting angle  approaches $\pm \pi/2$ as $\varphi_{ti}$ increases.
We can also express the tilting  angle $\theta$ in terms of the Fermi velocity $v_{F}$, defined as the positive slope of the spectral cone. See figure~\ref{figure9a}. 
The tilting  angle can also be given as a function of $v_{F}$: 
\begin{equation}
	\label{eq5.9}
	\theta = -\dfrac{\pi}{4} + \arctan v_{F}.
\end{equation}
We plot this relation in the figure~\ref{figure9c}.
As the Fermi velocity increases,  the spectral cone  asymptotically approaches to  the critically tilted
case. 	Finally, we find a relation between $\varphi_{ti}$ and the Fermi velocity by using \eqref{eq5.8} and \eqref{eq5.9} such that
\begin{equation}
	\label{eq5.10}
	\varphi_{ti} = \dfrac{v_{F} - 1}{v_{F} + 1},\quad\text{for}\,\,\, v_{F}\geq-1.
\end{equation}
We plot this relation in the figure~\ref{figure9d}.

So far, we have  analyzed the Green's function and spectral densities for one-flavor spinors under symmetric tensor coupling. 
In the next section, this framework is extended to incorporate two-flavor spinors.

\section{Two-flavor cases}
\label{section4}
In this section, we analyze the cases with two-flavor spinors coupled with symmetric tensor. 
The Green's function and spectral densities for two-flavor spinors can be derived directly from those of the  one-flavor cases  by  combining   positive and negative coupling   in the one-flavor cases. And so is the  classification. 

\subsection{Dirac equation and source identification}
\label{section4.1}
For simplicity, we assume the masses of the flavors are identical.
The actions for the two-flavor spinors  with  symmetric tensor coupling is given by   
\begin{gather}
	\label{eq4.2}
	S_{\mathrm{bulk}} = i\sum_{i = 1}^{2}\int_{\mathcal{M}}d^{5}x\sqrt{-g}\,\bar{\psi}^{(i)}(\overleftrightarrow{\slashed{D}} - m)\psi^{(i)},\\
	\label{eq4.3}
	\begin{aligned}
		S_{\mathrm{int}} = i\int_{\mathcal{M}}d^{5}x\sqrt{-g}\bigg(&\bar{\psi}^{(1)}\Phi_{MN}\Gamma^{(M}\overleftrightarrow{D}^{N)}\psi^{(2)}\\
		+ &\bar{\psi}^{(2)}\Phi_{MN}\Gamma^{(M}\overleftrightarrow{D}^{N)}\psi^{(1)}\bigg),
	\end{aligned}\\
	\label{eq4.4}
	S_{\mathrm{bdy}} = -i\sum_{i = 1}^{2}\int_{\partial\mathcal{M}}d^{4}x\sqrt{-gg^{rr}}\bar{\psi}^{(i)}\psi^{(i)}.
\end{gather}
From the   action, the bulk equations of motion for $\psi^{(i)}$ are given by
\begin{equation}
	\label{eq4.5}
	\begin{pmatrix}
		\slashed{D} - m & \Phi_{MN}\Gamma^{(M}D^{N)}\\
		\Phi_{MN}\Gamma^{(M}D^{N)} & \slashed{D} - m
	\end{pmatrix}\begin{pmatrix}
		\psi^{(1)}\\
		\psi^{(2)}
	\end{pmatrix} = 0.
\end{equation}
As in the one-flavor case, we adopt the following ansatz for the two-spinor field:
\begin{equation}
	\label{eq4.6}
	\psi^{(i)}
	= (-gg^{rr})^{-1/4}e^{-i\omega t + ik_{x}x + ik_{y}y + ik_{z}z} 
	\zeta^{(i)} .
\end{equation}

To define the source and condensation for two-flavor spinors, we decompose each flavor field as
\begin{equation}
	\psi^{(1)} = \begin{pmatrix}
		\psi^{(1)}_{+}\\
		\psi^{(1)}_{-}
	\end{pmatrix}  \quad\text{and}\quad
	\psi^{(2)} = \begin{pmatrix}
		\psi^{(2)}_{+}\\
		\psi^{(2)}_{-}
	\end{pmatrix} ,
\end{equation}
where each $\psi_{\pm}^{(i)}$ ($i = 1, 2$) is a two-component spinor field. From the field decomposition in \eqref{eq4.6}, we decompose $\zeta^{(i)}$ ($i = 1, 2$) in the same manner as
\begin{equation}
	\zeta^{(1)} = \begin{pmatrix}
		\zeta^{(1)}_{+}\\
		\zeta^{(1)}_{-}
	\end{pmatrix}  
	\quad\text{and}\quad
	\zeta^{(2)}= \begin{pmatrix}
		\zeta^{(2)}_{+}\\
		\zeta^{(2)}_{-}
	\end{pmatrix} .
\end{equation}
Then, the boundary action \eqref{eq4.4} can be rewritten as
\begin{equation}
	\label{eq4.9}
	\begin{aligned}
		S_{\mathrm{bdy}} &= -i\sum_{i = 1}^{2}\int_{\partial\mathcal{M}}d^{4}x\bar{\zeta}^{(i)}\zeta^{(i)}\\
		&=  -\sum_{i = 1}^{2}\int_{\partial\mathcal{M}}d^{4}x\,{\zeta^{\dagger}_{+}}^{(i)}\zeta_{-}^{(i)} + \mathrm{h.c.}
	\end{aligned}
\end{equation}
When we vary the bulk action with respect to $\psi^{(i)}$ and add the boundary action variation, one can find that the total action variation can be expressed only in terms of the variation of $\zeta^{(1)}_{+}$ and $\zeta^{(2)}_{+}$ if the equations of motion are satisfied. 
Therefore, for the standard-standard (SS) quantization method, we identify the source and condensation as the boundary quantities of $(\zeta^{(1)}_{+}, \zeta^{(2)}_{+})^T$ and $(\zeta^{(1)}_{-}, \zeta^{(2)}_{-})^T$, respectively.
For notational clarity, we denote these bulk quantities as
\begin{equation}
	\label{eq4.10}
	\xi^{(S)} \equiv \begin{pmatrix}
		\zeta^{(1)}_{+}\\
		\zeta^{(2)}_{+}
	\end{pmatrix}\quad\mathrm{and}\quad\xi^{(C)} \equiv \begin{pmatrix}
		\zeta^{(1)}_{-}\\
		\zeta^{(2)}_{-}
	\end{pmatrix}.
\end{equation}
To extract the source and condensation from $\xi^{(S)}$ and $\xi^{(C)}$, we examine the boundary behavior of $\xi^{(S)}$ and $\xi^{(C)}$ by solving the equations of motion in \eqref{eq4.5}.
We observe that the leading terms of $\xi^{(S)}$ and $\xi^{(C)}$ are identical to those of the one-flavor case, as expressed in \eqref{eq3.11}.
Thus, we use the same notation, $\mathcal{J}$ and $\mathcal{C}$, as in \eqref{eq3.11} for the source and condensation of two-flavor spinors, respectively.
We utilize this identification of source and condensation when defining the retarded Green’s function for two-flavor spinors.

\subsection{Green's function and spectral densities}
\label{section4.3}
To define the retarded Green's function for two-flavor spinors, we employ our previous development for one-flavor spinors in the section~\ref{section3.3}.

Because $\xi^{(S)}$ and $\xi^{(C)}$ has the same form as \eqref{eq3.11}, we adopt the same calculations in \eqref{eq3.12}–\eqref{eq3.16} to express $\mathcal{J}$ and $\mathcal{C}$ in terms of $\mathbb{S}_{0}$ and $\mathbb{C}_{0}$.
In addition, notice that the effective action can be expressed in the same form as \eqref{eq3.17} by rewriting the boundary action in terms of $\xi^{(S)}$ and $\xi^{(C)}$.
Therefore, we do the same calculations in \eqref{eq3.17}–\eqref{eq3.20} to define the Green’s function for two-flavor spinors, denoted as $\mathbb{G}_{R}$, which is expressed as\footnote{Here, we distinguish the notation for the retarded Green’s function of two-flavor spinors, denoted by $\mathbb{G}_{R}$, from that of one-flavor spinors, denoted by $G_{R}$.}
\begin{equation}
	\label{eq4.16}
	\mathbb{G}_{R} = \mathbb{C}_{0}\mathbb{S}_{0}^{-1}.
\end{equation}

\begin{figure*}[tb]
	\centering
	\includegraphics[width =0.8\linewidth]{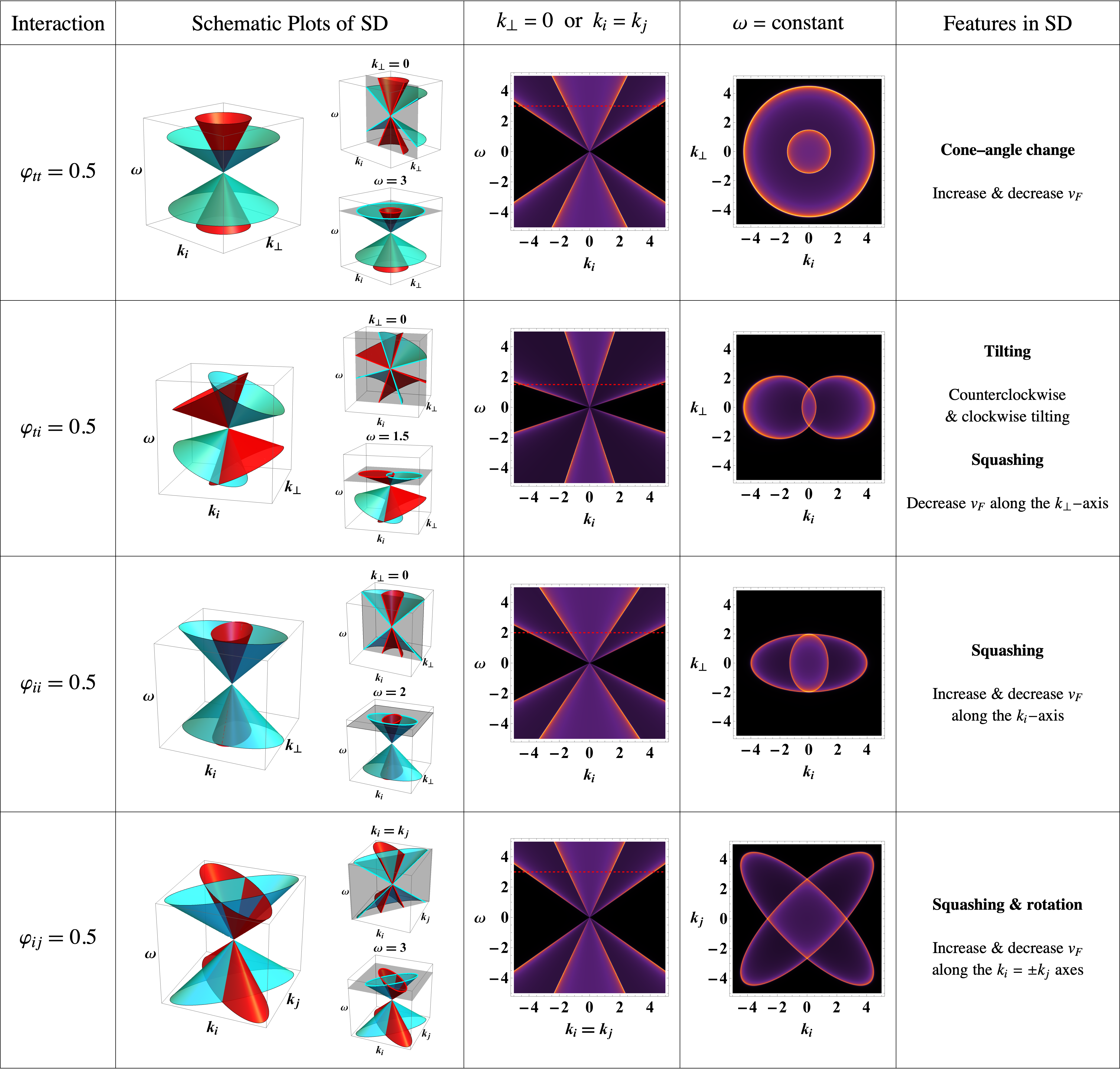}
	\caption{Classification of the spectral density (SD) for massless two-flavor fermions under each type of symmetric-tensor coupling. Schematic plots of the SD with $\varphi_{\mu\nu} = 0.5$ and their cross-sectional slices (gray planes and red dotted lines) are shown. Red cones represent contributions from $\Phi_{\mu\nu} > 0$, while blue ones represent contributions from $-\Phi_{\mu\nu} $.}
	\label{figure4}
\end{figure*}

We find that the Green's function in this case can be obtained from that of the  one-flavor cases in \eqref{eq3.23} via a similarity transformation, as detailed in Appendix~\ref{A3}. 
The result is given by
\begin{equation}
	\label{eq4.17}
	\mathbb{G}_{R} = \frac{1}{2}\begin{pmatrix}
		G_{R}(h) + G_{R}(-h) & & G_{R}(h) - G_{R}(-h) \\
		G_{R}(h) - G_{R}(-h) & & G_{R}(h) + G_{R}(-h)
	\end{pmatrix},
\end{equation}
where $G_{R}(h)$ is the Green's function for one-flavor spinors. 
Notice that we took Standard quantization for both fermion flavors.  
Now, the traced Green’s function for two-flavor spinors can be expressed in terms of the one-flavor Green's function: 
\begin{equation}
	\label{eq4.18}
	\Tr\mathbb{G}_{R} = \Tr G_{R}(h) + \Tr G_{R}(-h).
\end{equation}
Using the table~\ref{table2} and \eqref{eq4.18}, we get  the traced Green’s function    for  each tensor coupling and we present the result in the table~\ref{table3}. 
\begin{table*}\small
	\centering
	\resizebox{0.8\textwidth}{!}{
		\begin{tabular}[c]{| c | c |}
			\hline
			\textbf{Int.} & \textbf{Trace of analytic retarded Green's function} \\
			\hline
			$\Phi_{tt}$ & \begin{minipage}{0.9\textwidth}\vspace{5pt}
				\begin{equation}
					\Tr\mathbb{G}_{R} = \dfrac{2\omega}{\sqrt{\dfrac{\boldsymbol{k}^{2}}{(1 - \varphi_{tt})^{2}} - \omega^{2}}} + (\varphi_{tt}\,\leftrightarrow\,-\varphi_{tt})
				\end{equation}\vspace{0pt}
			\end{minipage} \\
			\hline
			$\Phi_{ti}$ & \begin{minipage}{0.9\textwidth}\vspace{5pt}
				\begin{equation}
					\Tr\mathbb{G}_{R} = \dfrac{2(\omega - \varphi_{ti} k_{i})}{\sqrt{
							\begin{pmatrix} \omega & k_{i} \end{pmatrix}
							\begin{pmatrix}
								-1 + \varphi_{ti}^{2} & 2 \varphi_{ti} \\
								2 \varphi_{ti} & 1 - \varphi_{ti}^{2}
							\end{pmatrix}
							\begin{pmatrix} \omega \\ k_{i} \end{pmatrix}
							+ |\boldsymbol{k}_{\perp}|^{2}}} + (\varphi_{ti}\,\leftrightarrow\,-\varphi_{ti})
				\end{equation}\vspace{0pt}
			\end{minipage} \\
			\hline
			$\Phi_{ii}$ & \begin{minipage}{0.9\textwidth}\vspace{5pt}
				\begin{equation}
					\Tr\mathbb{G}_{R} = \dfrac{2\omega}{\sqrt{(1 + \varphi_{ii})^{2} k_{i}^{2} + |\boldsymbol{k}_{\perp}|^{2} - \omega^{2}}} + (\varphi_{ii}\,\leftrightarrow\,-\varphi_{ii})
				\end{equation}\vspace{0pt}
			\end{minipage} \\
			\hline
			$\Phi_{ij}$ & \begin{minipage}{0.9\textwidth}\vspace{5pt}
				\begin{equation}
					\Tr\mathbb{G}_{R} = \dfrac{2\omega}{\sqrt{
							\begin{pmatrix} k_{i} & k_{j} \end{pmatrix}
							\begin{pmatrix}
								1 + \varphi_{ij}^{2} & 2 \varphi_{ij} \\
								2 \varphi_{ij} & 1 + \varphi_{ij}^{2}
							\end{pmatrix}
							\begin{pmatrix} k_{i} \\ k_{j} \end{pmatrix}
							+ k_{\perp}^{2} - \omega^{2}}} + (\varphi_{ij}\,\leftrightarrow\,-\varphi_{ij})
				\end{equation}\vspace{0pt}
			\end{minipage} \\
			\hline
		\end{tabular}
	}
	\caption{Traces of the analytic retarded Green’s function for two-flavor spinors under each interaction type (Int.) of symmetric tensor couplings.}
	\label{table3}
\end{table*}
We observe the same classification of interaction types as in the one-flavor case discussed in Section~\ref{section3.4}, based on the components $\Phi_{tt}$, $\Phi_{ti}$, $\Phi_{ii}$, and $\Phi_{ij}$.
By taking the imaginary part of \eqref{eq4.18}, we obtain the spectral densities in SS-quantization in terms of those for the one-flavor case, such that
\begin{equation}
	\label{eq4.23}
	\begin{aligned}
		A(\omega, \boldsymbol{k}) &= \mathrm{Im}\left(\Tr\mathbb{G}_{R}\right)\\
		&= \mathrm{Im}\left(\Tr G_{R}(h)\right) + \mathrm{Im}\left(\Tr G_{R}(-h)\right).
	\end{aligned}
\end{equation}
Thus, the spectral densities for two-flavor spinors are a combination of those for one-flavor cases with both positive and negative sign of interactions. 
From this,  we get  four types of spectral densities presented  in the figure~\ref{figure4}.
From this figure and table~\ref{table3}, we see that the cone-angle change, 
squashing, and tilting for the one-flavor case can be seen here for both positive and negative coupling.

\section{Holography of symmetric tensor coupling  vs.   Real material}
\label{section6}
In this section, we explore the application of our holographic mean-field theory with symmetric tensor coupling to condensed matter systems, focusing on the   features of spectral densities. 
To achieve this, we establish the correspondence between the parameters $\varphi_{\mu\nu}$ and their counterparts in condensed matter systems by focusing  on the local cone near the Dirac point.  

\begin{itemize}
	\item \textbf{$\Phi_{tt}$ vs. tuning Fermi velocity}\\
	As shown in the table~\ref{table2} and figure~\ref{figure3}, the $\Phi_{tt}$ component induces cone-angle change, leading to a uniform scaling of the Fermi velocity.
	The Fermi velocity of Dirac materials such as  graphene, can be fine-tuned by applying a uniform electric field, as illustrated in the figure~\ref{figure10}(a)  quoted from  \cite{D_az_Fern_ndez_2017}.
	\begin{figure*}[tb]
		\centering
		\includegraphics[width =0.75\linewidth]{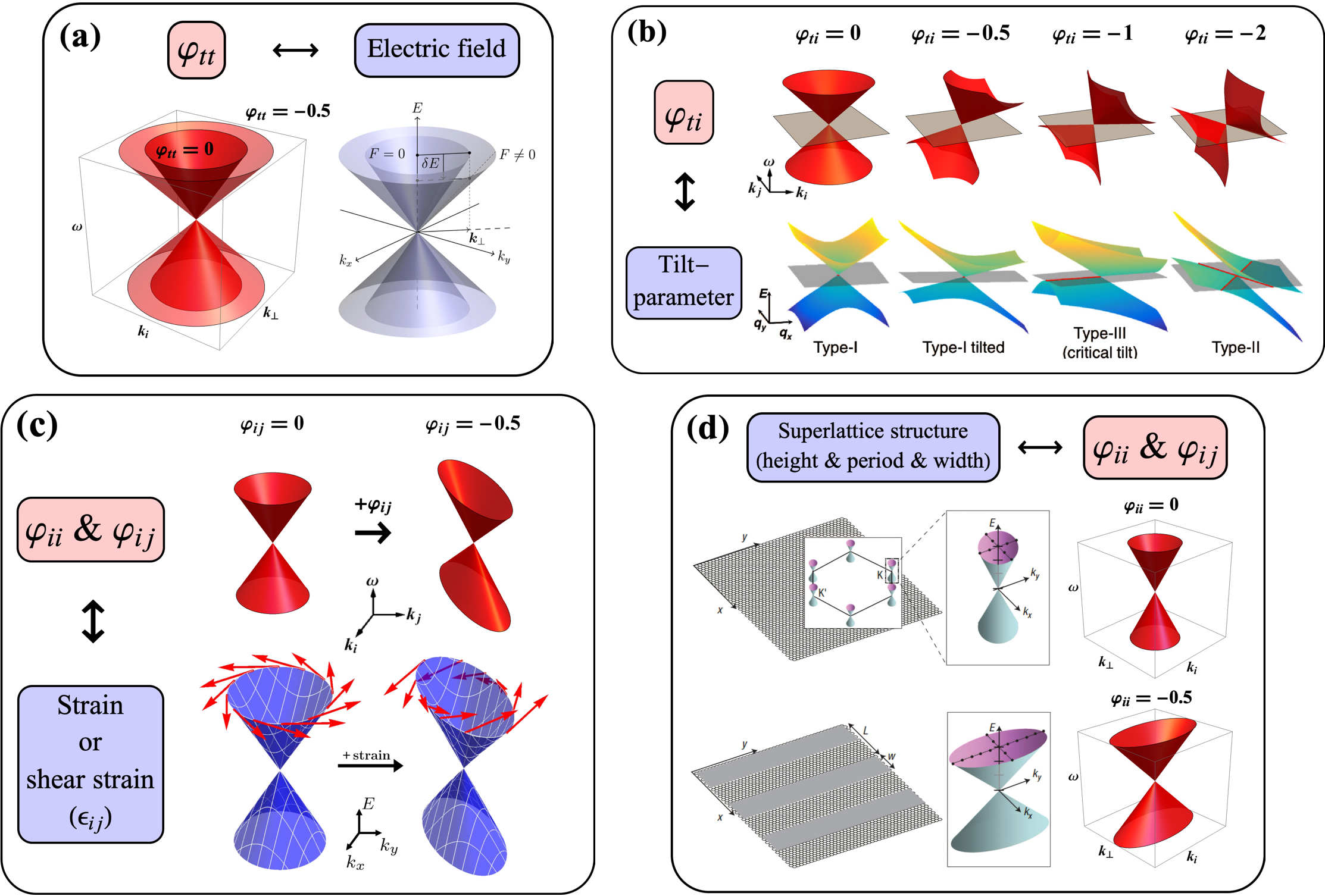}
		\caption{ Holography vs Real material : 
			The red   plots represent spectral densities  for our holographic model while the blue plots represent those in real material system. 
			(a) isotropic tuning of the Fermi velocity induced by a uniform electric field $F$, adapted from \cite{D_az_Fern_ndez_2017}; 
			(b) Tilted Dirac cones, adapted from \cite{PhysRevX.9.031010}; 
			(c) anisotropic Dirac cone on the surface of  $\mathrm{Bi}_{2}\mathrm{Se}_{3}$ $(111)$   under strain or shear strain, adapted from \cite{Brems_2018}; 
			(d)   squashed Dirac cones in superlattice graphene in periodic potential, adapted from \cite{Park_2008}.}
		\label{figure10}
	\end{figure*}
	Therefore, the value of $\varphi_{tt}$ serves as the effect of an electric field, enabling isotropic tuning of the Fermi velocity of Dirac cones.
	
	\item \textbf{$\Phi_{tt}$ vs. twist angle}\\
	The dispersion relation in \eqref{eq3.35} flattens as $\varphi_{tt} \to -\infty$, suggesting a significant suppression of the Fermi velocity. 
	Such locally flat energy spectra are observed in twisted bilayer graphene, where flat bands emerge at the magic angle \cite{Lisi_2020}. 
	This similarity gives the possible interpretation of $\varphi_{tt}$ as a parameter analogous to the twist angle in bilayer graphene.
	
	\item \textbf{$\Phi_{ti}$ vs. tilted cones}\\
	$\varphi_{ti}$ serves as a tilt parameter for tilted type-I, II, and III Dirac cones, or Weyl cones at single local Weyl nodes in Weyl semimetals \cite{Soluyanov_2015}, as illustrated in the figure~\ref{figure10}(b).
	
	\item \textbf{$\Phi_{ii}$ and $\Phi_{ij}$ vs. spin nematic order $Q_{ij}$}\\
	Previously, the   spatial symmetric tensor has been used in $d$-wave holographic superconductor.
	However, our model does not give gap since our model does not have complex  field coupled with $U(1)$ gauge field \cite{Kim:2013oba,PhysRevD.109.066004,Ghorai:2023wpu}. 
	%
	In the spin-nematic action within mean-field theory, the nematic order term \cite{PhysRevB.96.235140} parallels the interaction term in \eqref{eq2.5}. 
	So our model has more analogy with 
	the spin-nematic phase   characterized by a nematic order parameter $Q_{ij}$, a quadrupole rank-2 symmetric and traceless tensor \cite{PhysRevB.64.195109,PhysRevB.96.235140}. 
	Additionally, the spin nematic phase exhibits an anisotropic Fermi surface \cite{PhysRevB.96.235140}, which aligns with the squashing of spectral densities under $\Phi_{ii}$ and $\Phi_{ij}$ couplings shown in the figure~\ref{figure3}. 
	These parallels suggest that the spatial components $\varphi_{ii}$ and $\varphi_{ij}$ correspond to the  $Q_{ij}$ in the spin nematic phase. An analogy between the nematic order and gravitational waves is discussed in \cite{PhysRevB.109.L220407}.
	
	\item \textbf{$\Phi_{ii}$ and $\Phi_{ij}$  vs. strain tensor $\epsilon_{ij}$  }\\
	Squashing effect on spectral cone can  also be   found in two-dimensional graphene, where Dirac cones undergo anisotropic deformation due to strain or shear strain \cite{PhysRevB.88.085430,PhysRevResearch.4.L022027}.
	Additionally, as shown in the figure~\ref{figure10}(c), anisotropic Dirac cones with elliptical equi-energy contours emerge on the  $\mathrm{Bi}_{2}\mathrm{Se}_{3}$ $(111)$ surface under applied strain \cite{Brems_2018}.
	These deformations arise via a rank-2 strain tensor $\epsilon_{ij}$ in position or momentum space. 
	It is remarkable that the strain tensor can be quantitatively matched with $\varphi_{ii}$ and $\varphi_{ij}$, as detailed in Appendix~\ref{A4}.
	As a result, the values of $\varphi_{ii}$ and $\varphi_{ij}$ match with the strain tensor $\epsilon_{ij}$, describing Dirac cones that are squashed in the preferred direction.
	
	\item \textbf{$\Phi_{ii}$ and $\Phi_{ij}$   vs. superlattice structure}\\
	\sloppy
	In the superlattice graphene, Dirac cones exhibit anisotropic deformation along a preferred direction in momentum space due to a periodically distributed potential \cite{Park_2008}, as illustrated in the figure~\ref{figure10}(d). 
	As discussed in \cite{Park_2008}, the renormalized Fermi velocity depends on the height, period, and width of the periodic potential. 
	Therefore, the structural parameters of the potential can be corresponded to $\varphi_{ii}$ and $\varphi_{ij}$.
\end{itemize}

The symmetric tensor parameter $\varphi_{\mu\nu}$   can   be   matched with measured  parameters to  reproduce the local energy dispersion of a material. 
For example, the Dirac cone on the $\mathrm{Bi}_{2}\mathrm{Se}_{3}$ $(221)$ surface exhibits an anisotropic electronic band structure with elliptical equi-energy contours \cite{PhysRevB.84.195425}. 
This anisotropic feature corresponds to the squashed spectral density induced by   $\Phi_{ii}$. 
This correspondence enables  us to identify a negative value for $\varphi_{yy}$ to accurately reproduce the elliptical contours reported in \cite{PhysRevB.84.195425}.

	Although the deformed Green's functions can be expressed through linear transformations of frequency and momentum, these transformations are used only as a computational mapping to the free Green's function. 
	The physical spectral function is evaluated in the original variables $(\omega,k_i)$, which define the fixed laboratory frame. 
	In this frame, the external tensor source $\varphi_{\mu\nu}$ is varied while the lattice axes, strain directions, external-field directions, and measurement frame are held fixed. 
	Therefore, the resulting changes in cone angle, velocity, anisotropy, rotation, and tilt modify the measured dispersion relative to these fixed structures and cannot be removed by a passive coordinate redefinition. 
	We therefore interpret these effects as physically measurable spectral deformations induced by the external tensor source, rather than merely kinematic artifacts of the linear transformation.

In summary, the values of $\varphi_{\mu\nu}$ can be determined by measuring the Fermi velocities, cone angles, and tilt angles of the local energy dispersion for a given material (See figure~\ref{figure1}).

\section{\label{section7}Discussion}
In this work, we developed a holographic mean-field theory for one- and two-flavor spinors interacting with a rank-2 symmetric tensor field in an $\mathrm{AdS}_{5}$ background. 
By treating $\varphi_{\mu\nu}$ as a external source, we identified three primary effects on fermionic spectral densities: cone-angle change, squashing, and tilting, including over-tilted regimes.

For one-flavor spinors, we derived analytic solutions for the retarded Green’s function and spectral densities in pure $\mathrm{AdS}_{5}$, demonstrating how each boundary component $\Phi_{\mu\nu}$ deforms the shape, cone angle, and Fermi velocity of the spectral densities. 
Extending this framework, we constructed two-flavor retarded Green’s functions by combining single-flavor solutions with opposite coupling signs. 
These results enabled a systematic classification of the symmetric tensor field’s roles into cone-angle change, squashing, and tilting effects. 
We further proposed a method for constructing well-defined over-tilted spectral densities in the regime $|\varphi_{ti}| > 1$ by considering $i\epsilon$ prescription in a transformed momentum space to preserve the causality condition in a covariant way.

\sloppy
In the section~\ref{section6}, we explored physical applications of symmetric tensor couplings, showing a potential phenomenological connection between holography and condensed matter physics.
The $\varphi_{tt}$ component governs isotropic Fermi velocity rescaling, observed in Dirac materials under a uniform electric field. 
Indeed, strong $\varphi_{tt}$ coupling significantly flattens the dispersion, resembling the flat band observed in twisted bilayer graphene at the magic angle.
The $\varphi_{ti}$ components act as a tilt parameter for type-I Dirac or Weyl cones, while over-tilted spectral densities correspond to type-II or type-III Dirac/Weyl semimetals, known for their unique transport properties and causal structures. 
Spatial components $\varphi_{ii}$ and $\varphi_{ij}$ can be matched to quadrupolar nematic order in spin-nematic systems and strain tensors in strained graphene or $\mathrm{Bi}_{2}\mathrm{Se}_{3}$-class materials. 
Furthermore, the anisotropic spectral densities from these spatial components align with superlattice graphene under periodic potentials. 
These connections suggest that symmetric tensor couplings in our holographic model can qualitatively and quantitatively capture the spectral and energy dispersion features of real materials. 
This framework provides a holographic dictionary for phenomena such as tilted Dirac cones and lattice-driven anisotropic spectral properties.

We want to mention several directions for future research. 
First, we gave a qualitative mapping between symmetric tensor components and  nematic order.  
And it  would be interesting to  establish   more   quantitative correspondence. 
Second, exploring the topological properties induced by symmetric tensor components, such as Berry curvature will offer promising avenues for further study.

\section*{Acknowledgements}
	This work is supported by Mid-career Researcher Program through the National Research
	Foundation of Korea grant No. NRF-2021R1A2B5B02002603, RS-2023-00218998. 
	We thank the APCTP for the hospitality during the focus program, where part of this work
	was discussed.

\appendix
\section{Flow equation: derivation and analytical solutions}
\label{A1}
In this section, we explicitly derive the flow equation in \eqref{eq3.24} and solve it to obtain the analytic bulk quantity $\mathbb{G}(r)$ in \eqref{eq3.29}.

To set up the flow equation, we substitute the expression in \eqref{eq3.12} and \eqref{eq3.13} into the equations of motion in \eqref{eq3.9} and \eqref{eq3.10}.
Then, we get
\begin{gather}
	\label{eqA.1}
	\partial_{r}\mathbb{S}(r) + \mathbb{M}_{1}\mathbb{S}(r) + \mathbb{M}_{2}\mathbb{C}(r) = 0,\\
	\label{eqA.2}
	\partial_{r}\mathbb{C}(r) + \mathbb{M}_{3}\mathbb{C}(r) + \mathbb{M}_{4}\mathbb{S}(r) = 0,
\end{gather}
because ${\bf c}$ is an arbitrary vector in the solution space. 
When we take a derivative with respect to $r$ on the bulk quantity $\mathbb{G}(r)$ in \eqref{eq3.21}, since $\partial_{r}\mathbb{S}^{-1} = -\mathbb{S}(r)^{-1}(\partial_{r}\mathbb{S}(r))\mathbb{S}^{-1}$, we have
\begin{equation}
	\label{eqA.3}
	\begin{aligned}
		\partial_{r}\mathbb{G} &= \partial_{r}(\mathbb{C}(r)\mathbb{S}(r)^{-1})\\
		&= (\partial_{r}\mathbb{C}(r))\mathbb{S}(r)^{-1} + \mathbb{C}(r)\partial_{r}\mathbb{S}(r)^{-1}\\
		&= (\partial_{r}\mathbb{C}(r))\mathbb{S}(r)^{-1} - \mathbb{C}(r)\mathbb{S}(r)^{-1}(\partial_{r}\mathbb{S}(r))\mathbb{S}(r)^{-1}.
	\end{aligned}
\end{equation}
If we substitute \eqref{eqA.1} and \eqref{eqA.2} for $\partial_{r}\mathbb{S}(r)$ and $\partial_{r}\mathbb{C}(r)$ in \eqref{eqA.3}, above equation becomes
\begin{equation}
	\label{eqA.4}
	\partial_{r}\mathbb{G} = \mathbb{G}(r)\mathbb{M}_{2}\mathbb{G}(r) + \mathbb{G}(r)\mathbb{M}_{1} - \mathbb{M}_{3}\mathbb{G}(r) - \mathbb{M}_{4}.
\end{equation}
By defining $\bar{\mathbb{M}}_{1} \equiv -\mathbb{M}_{1}$ and $\bar{\mathbb{M}}_{2} \equiv -\mathbb{M}_{2}$, we arrive at the desired flow equation presented in \eqref{eq3.24}.
For the metric in \eqref{eq3.2}, the four matrices in the flow equation are given by
\begin{equation}
	\label{eqA.6}
	\begin{gathered}
		\begin{aligned}
			\bar{\mathbb{M}}_{1} &= \left(\frac{m}{r\sqrt{f}} + \frac{\Tr\Phi/2}{f} + \frac{r^{2}f^{\prime}}{4f^{2}}\Phi_{tt}\right)\mathbbm{1}_{2\times2} + \frac{r^{2}f^{\prime}}{4f\sqrt{f}}\Phi_{ti}\sigma^{i},\\
			\bar{\mathbb{M}}_{2} &= -\dfrac{r^{2}}{f}\left[(g + \Phi)_{t\mu}\bar{k}^{\mu}\right]\mathbbm{1}_{2\times2} - \dfrac{r^{2}}{\sqrt{f}}(g + \Phi)_{\mu i}\bar{k}^{\mu}\sigma^{i},\\
			\mathbb{M}_{3} &= \left(\frac{m}{r\sqrt{f}} - \frac{\Tr\Phi/2}{f} - \frac{r^{2}f^{\prime}}{4f^{2}}\Phi_{tt}\right)\mathbbm{1}_{2\times2} + \frac{r^{2}f^{\prime}}{4f\sqrt{f}}\Phi_{ti}\sigma^{i},\\
			\mathbb{M}_{4} &= -\dfrac{r^{2}}{f}\left[(g + \Phi)_{t\mu}\bar{k}^{\mu}\right]\mathbbm{1}_{2\times2} + \dfrac{r^{2}}{\sqrt{f}}(g + \Phi)_{\mu i}\bar{k}^{\mu}\sigma^{i},
		\end{aligned}\\
		\text{where}\quad\bar{k}^{\mu}\equiv(\omega/f, \boldsymbol{k}).
	\end{gathered}
\end{equation}
By letting $f = 1$ for the pure $\mathrm{AdS}_{5}$ geometry, we can recover the four matrices in \eqref{eq3.26}.

Now, we explicitly show how to derive the solution for $\mathbb{G}(r)$ in \eqref{eq3.29} using a specific ansatz for it.
To solve the flow equation in \eqref{eq3.27}, we set an ansatz of $\mathbb{G}(r)$ as
\begin{equation}
	\label{eqA.7}
	\mathbb{G}(r) = \begin{pmatrix}
		\mathcal{G}_{11}(r) & \mathcal{G}_{12}(r)\\
		\mathcal{G}_{21}(r) & \mathcal{G}_{22}(r)
	\end{pmatrix} = \begin{pmatrix}
		a_{11} & a_{12}\\
		a_{21} & a_{22}
	\end{pmatrix}\mathcal{G}(r) = \mathbb{A}\mathcal{G}(r),
\end{equation}
where $a_{ij}$ are some constants while $\mathcal{G}(r)$ is a function. 
We substitute this ansatz into the flow equation \eqref{eq3.27}, so it becomes
\begin{equation}
	\label{eqA.8}
	\begin{aligned}
		\mathbb{A}\partial_{r}\mathcal{G}(r) &- r^{2}\mathbb{A}\sigma^{\mu}\mathbb{A}(g + \Phi)_{\mu\nu}k^{\nu}\mathcal{G}^{2}(r) + \frac{2m}{r}\mathbb{A}\mathcal{G}(r)\\
		&- r^{2}\bar{\sigma}^{\mu}(g + \Phi)_{\mu\nu}k^{\nu} = 0.
	\end{aligned}
\end{equation}
We normalize the coefficients ahead of $\partial_{r}\mathcal{G}$ by dividing $a_{11}$, $a_{12}$, $a_{21}$, and $a_{22}$ on each element of \eqref{eqA.8}. 
That is, we multiply a matrix ${\scriptsize\begin{pmatrix}
		1 & 1\\
		1 & 1
\end{pmatrix}}\mathbb{A}^{-1}$ on \eqref{eqA.8} so that
\begin{equation}
	\label{eqA.9}
	\begin{aligned}
		\begin{pmatrix}
			1 & 1\\
			1 & 1
		\end{pmatrix}\bigg(&\mathbbm{1}_{2\times2}\partial_{r}\mathcal{G}(r) - r^{2}\sigma^{\mu}\mathbb{A}(g + \Phi)_{\mu\nu}k^{\nu}\mathcal{G}^{2}(r)\\
		&+ \frac{2m}{r}\mathcal{G}(r) - r^{2}\mathbb{A}^{-1}\bar{\sigma}^{\mu}(g + \Phi)_{\mu\nu}k^{\nu}\bigg) = 0.
	\end{aligned}
\end{equation}
Then, we get four differential equations with different coefficients ahead of $\mathcal{G}^{2}(r)$ and the constant terms in \eqref{eqA.9}. We demand that these coefficients have to be same each other for a unique solution of $\mathcal{G}(r)$. 
Therefore, we impose
\begin{equation}
	\label{eqA.10}
	\begin{aligned}
		\sigma^{\mu}\mathbb{A}(g + \Phi)_{\mu\nu}k^{\nu} &= c_{1}\mathbbm{1}_{2\times2},\\
		\mathbb{A}^{-1}\bar{\sigma}^{\mu}(g + \Phi)_{\mu\nu}k^{\nu} &= c_{2}\mathbbm{1}_{2\times2},
	\end{aligned}
\end{equation}
where $c_{1}$ and $c_{2}$ are some constants. By solving these two relations simultaneously with respect to $c_{1}$, $c_{2}$, $a_{12}$, $a_{21}$, and $a_{22}$, we represent $a_{12}$, $a_{21}$, $a_{22}$ in terms of $a_{11}$ as
\begin{equation}
	\label{eqA.11}
	\begin{aligned}
		a_{12} &= -\frac{(\eta + \varphi)_{x\mu}k^{\mu} - i(\eta + \varphi)_{y\nu}k^{\nu}}{(\eta + \varphi)_{t\mu}k^{\mu} - (\eta + \varphi)_{z\mu}k^{\mu}}a_{11},\\
		a_{21} &= -\frac{(\eta + \varphi)_{x\mu}k^{\mu} + i(\eta + \varphi)_{y\nu}k^{\nu}}{(\eta + \varphi)_{t\mu}k^{\mu} - (\eta + \varphi)_{z\mu}k^{\mu}}a_{11},\\
		a_{22} &= \frac{(\eta + \varphi)_{t\mu}k^{\mu} + (\eta + \varphi)_{z\nu}k^{\nu}}{(\eta + \varphi)_{t\mu}k^{\mu} - (\eta + \varphi)_{z\mu}k^{\mu}}a_{11}.
	\end{aligned}
\end{equation}
Or, we rewrite these relations in the matrix representation as
\begin{equation}
	\begin{pmatrix}
		\label{eqA.12}
		a_{11} & a_{12}\\
		a_{21} & a_{22}
	\end{pmatrix} = \frac{\bar{\sigma}^{\mu}(\eta + \varphi)_{\mu\nu}k^{\nu}}{[(\eta + \varphi)_{t\rho}- (\eta + \varphi)_{z\rho}]k^{\rho}}a_{11}.
\end{equation}
By substituting this relation into \eqref{eqA.7}, we find
\begin{equation}
	\label{eqA.13}
	\begin{aligned}
		\mathbb{G}(r) &= \begin{pmatrix}
			\mathcal{G}_{11}(r) & \mathcal{G}_{12}(r)\\
			\mathcal{G}_{21}(r) & \mathcal{G}_{22}(r)
		\end{pmatrix} = \frac{1}{a_{11}}\begin{pmatrix}
			a_{11} & a_{12}\\
			a_{21} & a_{22}
		\end{pmatrix}\mathcal{G}_{11}(r)\\
		 &= \frac{\bar{\sigma}^{\mu}(\eta + \varphi)_{\mu\nu}k^{\nu}}{[(\eta + \varphi)_{t\rho}- (\eta + \varphi)_{z\rho}]k^{\rho}}\mathcal{G}_{11}(r).
	\end{aligned}
\end{equation}
Therefore, once we solve $\mathcal{G}_{11}(r)$, we solve $\mathbb{G}(r)$ as well. 
By putting \eqref{eqA.13} into \eqref{eq3.27}, we get four identical flow equations in the form of
\begin{equation}
	\label{eqA.14}
	\begin{gathered}
		\begin{aligned}
			\partial_{r}\mathcal{G}_{11}(r) &+ \frac{|(g + \Phi)_{\mu\nu}k^{\nu}|^{2}}{(g + \Phi)_{t\rho}k^{\rho}}r^{2}\mathcal{G}_{11}(r)^{2} + \frac{2m}{r}\mathcal{G}_{11}(r) \\
			&- r^{2}(g + \Phi)_{t\sigma}k^{\sigma} = 0,
		\end{aligned}\\
		\text{where}\quad|(g + \Phi)_{\mu\nu}k^{\nu}|^{2} = g^{\mu\nu}(g + \Phi)_{\mu\rho}k^{\rho}(g + \Phi)_{\nu\sigma}k^{\sigma}.
	\end{gathered}
\end{equation}
Also, from \eqref{eq3.28} and \eqref{eqA.7}, the IR boundary condition for $\mathcal{G}_{11}(r)$ is given by
\begin{equation}
	\label{eqA.15}
	\mathcal{G}_{11}(r\to\infty) = i.
\end{equation}
We obtain $\mathcal{G}_{11}(r)$ by solving \eqref{eqA.14} with the condition \eqref{eqA.15}, then substitute this solution into \eqref{eqA.13} to obtain $\mathbb{G}(r)$;
the resulting solution is given by \eqref{eq3.29}.

\section{Derivation of correlator for two-flavor spinors}
\label{A3}
In this section, we show how to derive the retarded Green's function for two-flavor spinors in \eqref{eq4.17}.

We begin with the bulk and interaction actions in \eqref{eq4.2} and \eqref{eq4.3}:
\begin{equation}
	\label{eqB.1}
	\begin{aligned}
		S_{\mathrm{bulk}} + S_{\mathrm{int}} = &i\int_{\mathcal{M}}d^{5}x\sqrt{-g}\begin{pmatrix}
			\bar{\psi}^{(1)} & \bar{\psi}^{(2)}
		\end{pmatrix}\\
		&\times\begin{pmatrix}
			\overleftrightarrow{\slashed{D}} - m & \Phi_{MN}\Gamma^{(M}\overleftrightarrow{D}^{N)}\\[5pt]
			\Phi_{MN}\Gamma^{(M}\overleftrightarrow{D}^{N)} & \overleftrightarrow{\slashed{D}} - m
		\end{pmatrix}\begin{pmatrix}
			\psi^{(1)}\\
			\psi^{(2)}
		\end{pmatrix}.
	\end{aligned}
\end{equation}
Using the field ansatz in \eqref{eq4.6}, we rewrite this action as
\begin{equation}
	\label{eqB.2}
	\begin{aligned}
		S_{\mathrm{bulk}} + S_{\mathrm{int}} = &i\int_{\mathcal{M}}d^{5}x\begin{pmatrix}
			\bar{\zeta}^{(1)} & \bar{\zeta}^{(2)}
		\end{pmatrix}\\
		&\times\begin{pmatrix}
			\overleftrightarrow{\slashed{D}}- m & \Phi_{MN}\Gamma^{(M}\overleftrightarrow{D}^{N)}\\[5pt]
			\Phi_{MN}\Gamma^{(M}\overleftrightarrow{D}^{N)} & \overleftrightarrow{\slashed{D}} - m
		\end{pmatrix}\begin{pmatrix}
			\zeta^{(1)}\\
			\zeta^{(2)}
		\end{pmatrix}.
	\end{aligned}
\end{equation}
We diagonalize the interaction term in \eqref{eqB.2} using a similarity transformation with an $8 \times 8$ matrix $\mathbb{U}$, given by
\begin{equation}
	\label{eqB.3}
	\mathbb{U} = \dfrac{1}{\sqrt{2}}\begin{pmatrix}
		\mathbbm{1}_{4\times4} & \mathbbm{1}_{4\times4}\\
		\mathbbm{1}_{4\times4} & -\mathbbm{1}_{4\times4}
	\end{pmatrix}\quad\text{so that}\,\,\,\mathbb{U}^{-1} = \mathbb{U}.
\end{equation}
The diagonalization results in
\begin{equation}
	\begin{adjustbox}{max width=\columnwidth}
			\label{eqB.4}
			$
			\begin{gathered}
			\begin{aligned}
				S&_{\mathrm{bulk}} + S_{\mathrm{int}} \\
				= &\, i\int_{\mathcal{M}}d^{5}x\begin{pmatrix}
					\bar{\zeta}^{(1)} & \bar{\zeta}^{(2)}
				\end{pmatrix}\mathbb{U}^{-1}\mathbb{U}\\
				&\times\begin{pmatrix}
					\overleftrightarrow{\slashed{D}} - m & \Phi_{MN}\Gamma^{(M}\overleftrightarrow{D}^{N)}\\[5pt]
					\Phi_{MN}\Gamma^{(M}\overleftrightarrow{D}^{N)} & \overleftrightarrow{\slashed{D}} - m
				\end{pmatrix}\mathbb{U}^{-1}\mathbb{U}\begin{pmatrix}
					\zeta^{(1)}\\
					\zeta^{(2)}
				\end{pmatrix}\\
				= &\, i\int_{\mathcal{M}}d^{5}x\begin{pmatrix}
					\bar{\phi}^{(1)} & \bar{\phi}^{(2)}
				\end{pmatrix}\\
				&\times\begin{pmatrix}
					\overleftrightarrow{\slashed{D}} - m + \Phi_{MN}\Gamma^{(M}\overleftrightarrow{D}^{N)} & 0\\
					0 & \overleftrightarrow{\slashed{D}} - m - \Phi_{MN}\Gamma^{(M}\overleftrightarrow{D}^{N)}
				\end{pmatrix}\begin{pmatrix}
					\phi^{(1)}\\
					\phi^{(2)}
				\end{pmatrix},
			\end{aligned}\\
			\begin{aligned}
				\mathrm{where}\quad(\phi^{(1)} , \phi^{(2)})^{T} &= (\phi^{(1)}_{+}, \phi^{(1)}_{-}, \phi^{(2)}_{+}, \phi^{(2)}_{-})^{T} = \mathbb{U}(\zeta^{(1)} , \zeta^{(2)})^{T}\\
				&= \mathbb{U}(\zeta^{(1)}_{+}, \zeta^{(1)}_{-}, \zeta^{(2)}_{+}, \zeta^{(2)}_{-})^{T}.
			\end{aligned}
			\end{gathered}
			$
	\end{adjustbox}
\end{equation}
Notice that the transformed fields $\phi^{(i)}$ ($i = 1, 2$) can be expressed in terms of $\xi^{(S)}$ and $\xi^{(C)}$ in \eqref{eq4.10} as
\begin{equation}
	\label{eqB.5}
	\begin{gathered}
		\begin{aligned}
			\begin{pmatrix}
				\phi^{(1)}_{+}\\
				\phi^{(2)}_{+}
			\end{pmatrix} &= \mathcal{U}\begin{pmatrix}
				\zeta^{(1)}_{+}\\
				\zeta^{(2)}_{+}
			\end{pmatrix} = \mathcal{U}\xi^{(S)},\\
			\begin{pmatrix}
				\phi^{(1)}_{-}\\
				\phi^{(2)}_{-}
			\end{pmatrix} &= \mathcal{U}\begin{pmatrix}
				\zeta^{(1)}_{-}\\
				\zeta^{(2)}_{-}
			\end{pmatrix} = \mathcal{U}\xi^{(C)},
		\end{aligned}\\
		\mathrm{where}\quad\mathcal{U} = \dfrac{1}{\sqrt{2}}\begin{pmatrix}
			\mathbbm{1}_{2\times2} & \mathbbm{1}_{2\times2}\\
			\mathbbm{1}_{2\times2} & -\mathbbm{1}_{2\times2}
		\end{pmatrix}\quad\text{so that}\,\,\,\mathcal{U}^{-1} = \mathcal{U}.
	\end{gathered}
\end{equation}
From \eqref{eqB.5}, the variation with respect to $\xi^{(S)}$ can be identified as the variation of the transformed field $(\phi_{+}^{(1)}, \phi_{+}^{(2)})^{T}$.
Likewise, the same identification holds for $\xi^{(C)}$ and $(\phi_{-}^{(1)}, \allowbreak \phi_{-}^{(2)})^{T}$.
Thus, the source for $\phi^{(i)}$ can be expressed in terms of that for $\psi^{(i)}$, and so can the condensation.
By observing the boundary behavior of $\phi^{(i)}$, this correspondence is
\begin{equation}
	\label{eqB.6}
	\begin{gathered}
		\begin{aligned}
			\begin{pmatrix}
				\phi^{(1)}_{+}\\
				\phi^{(2)}_{+}
			\end{pmatrix} &\approx r^{m + \Tr\varphi/2}\mathcal{U}\mathcal{J} = r^{m + \Tr\varphi/2}\mathcal{J}_{\phi},\\
			\begin{pmatrix}
				\phi^{(1)}_{-}\\
				\phi^{(2)}_{-}
			\end{pmatrix} &\approx r^{-m + \Tr\varphi/2}\mathcal{U}\mathcal{C} = r^{-m + \Tr\varphi/2}\mathcal{C}_{\phi},
		\end{aligned}\\
		\mathrm{where}\quad\mathcal{J}_{\phi} = \mathcal{U}\mathcal{J}\quad\mathrm{and}\quad\mathcal{C}_{\phi} = \mathcal{U}\mathcal{C}.
	\end{gathered}
\end{equation}
Consequently, we choose $\mathcal{J}_{\phi}$ as the source and $\mathcal{C}_{\phi}$ as the condensation for $\phi^{(i)}$.

Because $\phi^{(1)}$ and $\phi^{(2)}$ are decoupled in $\eqref{eqB.4}$, finding the Green’s function for $\phi^{(i)}$ is more straightforward than for $\psi^{(i)}$.
Thus, we first find the Green's function for $\phi^{(i)}$, denoted as $\mathbb{G}_{\phi}$.
We observe that the form of effective action in \eqref{eq4.9} remains identical for $\zeta^{(i)}$ and $\phi^{(i)}$, such that
\begin{equation}
	\label{eqB.7}
	S_{\mathrm{eff}} = -i\sum_{i = 1}^{2}\int_{\partial\mathcal{M}}d^{4}x \,\bar{\zeta}^{(i)}\zeta^{(i)} = -i\sum_{i = 1}^{2}\int_{\partial\mathcal{M}}d^{4}x \,\bar{\phi}^{(i)}\phi^{(i)}.
\end{equation}
Therefore, this effective action defines the same form of the retarded Green’s function for $\mathcal{J}$ and $\mathcal{J}_{\phi}$, such that
\begin{equation}
	\label{eqB.8}
	\begin{aligned}
		S_{\mathrm{eff}} &= -\int_{\partial\mathcal{M}}d^{4}x\,\epsilon^{\Tr\varphi}\mathcal{J}^{\dagger}\mathbb{G}_{R}\mathcal{J} + \mathrm{h.c.}\\
		&= -\int_{\partial\mathcal{M}}d^{4}x\,\epsilon^{\Tr\varphi}\mathcal{J}^{\dagger}_{\phi}\mathbb{G}_{\phi}\mathcal{J}_{\phi} + \mathrm{h.c.}
	\end{aligned}
\end{equation}
where $\mathbb{G}_{R}$ is given in \eqref{eq4.16}. 
By applying the transformation of source in \eqref{eqB.6} to \eqref{eqB.8}, $\mathbb{G}_{\phi}$ can be expressed in terms of $\mathbb{G}_{R}$ as
\begin{equation}
	\label{eqB.9}
	\mathbb{G}_{\phi} = \mathcal{U}\mathbb{G}_{R}\,\mathcal{U}.
\end{equation}
On the other hand, $\mathbb{G}_{\phi}$ is derived from the diagonalized action in \eqref{eqB.4} and is given by
\begin{equation}
	\label{eqB.10}
	\mathbb{G}_{\phi} = \mathrm{diag}\left(G_{R}(h), G_{R}(-h)\right),
\end{equation}
where $G_{R}(h)$ is the retarded Green’s function for a one-flavor spinor.
By using \eqref{eqB.9} and \eqref{eqB.10}, we find the retarded Green’s function for $\psi^{(i)}$ in terms of that for one-flavor spinors as
\begin{equation}
	\label{eqB.11}
	\begin{aligned}
		\mathbb{G}_{R} &= \mathcal{U}^{-1}\mathbb{G}_{\phi}\,\mathcal{U}^{-1}\\ 
		&= \frac{1}{2}\begin{pmatrix}
			\mathbbm{1}_{2\times2} & \mathbbm{1}_{2\times2}\\
			\mathbbm{1}_{2\times2} & -\mathbbm{1}_{2\times2}
		\end{pmatrix}\begin{pmatrix}
			G_{R}(h) & 0\\
			0 & G_{R}(-h)
		\end{pmatrix}\begin{pmatrix}
			\mathbbm{1}_{2\times2} & \mathbbm{1}_{2\times2}\\
			\mathbbm{1}_{2\times2} & -\mathbbm{1}_{2\times2}
		\end{pmatrix}\\
		&= \frac{1}{2}\begin{pmatrix}
			G_{R}(h) + G_{R}(-h) & & G_{R}(h) - G_{R}(-h)\\
			G_{R}(h) - G_{R}(-h) & & G_{R}(h) + G_{R}(-h)
		\end{pmatrix},
	\end{aligned}
\end{equation}
which is our previous result in \eqref{eq4.17}.

\section{Holographic tensor coupling vs. Strain tensor}
\label{A4}
As we explained in the section~\ref{section6}, the spatial components of the symmetric tensor correspond to the strain tensor, observed in strained two-dimensional graphene, for instance.
In this section, we quantitatively match the spatial components $\varphi_{ii}$ and $\varphi_{ij}$ with the strain tensor components $\epsilon_{ii}$ and $\epsilon_{ij}$.
We follow the explicit construction of the Hamiltonian for strained graphene in \cite{PhysRevB.88.085430} and compare its energy dispersion with that of our spectral density.
For notational clarity, we denote the indices $a, b = x, y$ for the strain tensor and spatial symmetric tensor components throughout this section.

We introduce a rank-2 symmetric strain tensor $\epsilon_{ab}$ as
\begin{equation}
	\label{C1}
	\epsilon_{ab} = \begin{pmatrix}
		\epsilon_{xx} & \epsilon_{xy}\\
		\epsilon_{xy} & \epsilon_{yy}
	\end{pmatrix},
\end{equation}
which changes the position of an atom as $\boldsymbol{x}\to(\mathbbm{1}_{2\times2} + \epsilon_{ab})\cdot\boldsymbol{x}$, where $\boldsymbol{x}$ is the two-dimensional position vector.
This strain tensor induces distortion of the reciprocal lattice, leading to anisotropic deformation of the Dirac cone.
Around the Dirac point, the Hamiltonian model is given by\cite{PhysRevB.88.085430}
\begin{equation}
	\label{C2}
	H = v_{0}\sigma^{a}(\delta_{ab} + (1 - \beta)\epsilon_{ab})q^{b},
\end{equation}
where $v_{0}$ is the Fermi velocity for the undeformed Dirac cone, $\sigma^{a} = (\sigma_{1}, \sigma_{2})$, $\beta$ is the difference in hopping energy between the deformed and undeformed lattice, and $q^{a}$ is the momentum measured relative to the Dirac points.
To compare this model with our spectral densities, we simply let $v_{0} = 1$ and $q^{a} = (k_{x}, k_{y})$ at the Dirac point. 
For the hypothetical case with $\beta = 0$, the Hamiltonian in \eqref{C2} reduces to
\begin{equation}
	\label{C3}
	H = v_{0}\sigma^{a}(\delta_{ab} + \epsilon_{ab})q^{b}.
\end{equation}
This $\beta = 0$ limit simplifies the analysis of lattice distortion effects by isolating them from hopping energy modifications.
The eigenvalues $E_{\pm}$ of the Hamiltonian in \eqref{C3} are
	\begin{equation}
	\label{C4}
	\begin{gathered}
		E_{\pm} = \pm\left[((1 + \epsilon_{xx})^{2} + \epsilon_{xy}^{2})k_{x}^{2} + 2\epsilon_{xy}(2 + \epsilon_{xx} + \epsilon_{yy})k_{x}k_{y}\right.\\
		\left.+ ((1 + \epsilon_{yy})^{2} + \epsilon_{xy}^{2})k_{y}^{2}\right]^{1/2}.
	\end{gathered}
\end{equation}
In our holographic model, the spectral density for $\Phi_{xx}$, $\Phi_{yy}$, and $\Phi_{xy}$ couplings can be derived from \eqref{eq3.31}.
For the comparison with the two-dimensional strained graphene, we let $k_{z} = 0$ so that the pole structure in $\Tr G_{R}$ is
	\begin{equation}
		\label{C6}
		\begin{gathered}
			\omega = \pm\left[((1 + \varphi_{xx})^{2} + \varphi_{xy}^{2})k_{x}^{2} + 2\varphi_{xy}(2 + \varphi_{xx} + \varphi_{yy})k_{x}k_{y}\right.\\
			\left.+ ((1 + \varphi_{yy})^{2} + \varphi_{xy}^{2})k_{y}^{2}\right]^{1/2},
		\end{gathered}
	\end{equation}
which can be exactly compared with the energy dispersion in \eqref{C4} by letting $\varphi_{ab}\leftrightarrow\epsilon_{ab}$.

If lattice deformation induces a change in hopping energy ($\beta \neq 0$), the identification is given by
\begin{equation}
	\varphi_{ab} \,\leftrightarrow\, (1 - \beta)\epsilon_{ab}, \quad \text{where} \quad a, b = x, y.
\end{equation}
In the realistic case with $\beta \approx 3$ \cite{PhysRevB.88.085430}, the sign of $\varphi_{ab}$ corresponding to $(1 - \beta)\epsilon_{ab}$ is flipped compared to the hypothetical case with $\beta = 0$. 
Physically, the flipping of $\varphi_{ab}$’s sign corresponds to a reversal in the squashing behavior of the spectral densities, consistent with the $\pi/2$ rotation of strain effects observed in graphene \cite{PhysRevB.88.085430} under the hopping-energy modifications.

\bibliographystyle{jhep}
\bibliography{TBH_ref.bib}

\end{document}